\documentclass[%
  reprint,
  superscriptaddress,
  nofootinbib,
]{revtex4-2}

\usepackage{amsmath}
\usepackage{amssymb}
\usepackage{bm,bbm}

\usepackage{graphicx}
\graphicspath{{figures/}}

\usepackage{enumitem}

\usepackage[bookmarks=false]{hyperref}
\hypersetup{colorlinks=true,citecolor=blue,linkcolor=red,%
urlcolor=blue,pdfstartview=FitH,bookmarksopen=true}

\usepackage[T1]{fontenc}
\usepackage[osf,sc]{mathpazo}
\usepackage{dsfont}

\usepackage{amsthm}
\newtheoremstyle{note}
{3pt}
{3pt}
{}
{}
{\itshape}
{:}
{.5em}
{}

\DeclareMathOperator{\Tr}{Tr}

\DeclareMathOperator{\id}{id}
\newcommand{\bra}[1]{\langle #1\rvert}
\newcommand{\ket}[1]{\lvert #1\rangle}
\newcommand{\ketbra}[1]{\lvert #1\rangle\langle #1\rvert}
\newcommand{\mean}[1]{\langle #1\rangle}

\newcommand{\braket}[2]{\langle #1\vert #2\rangle}
\newcommand{\abs}[1]{\lvert #1\rvert}

\newcommand{\md}{\mathrm{d}}
\newcommand{\me}{\mathrm{e}}
\newcommand{\mi}{\mathrm{i}}

\newcommand{\I}{\mathds{1}}
\newcommand{\Exp}[1]{\mathrm{e}^{#1}}

\newcommand{\maxover}[1][]{\underset{#1}{\mathrm{max}}}
\newcommand{\subto}{\mathrm{~s.t.}}
\newcommand{\FA}{\mathrm{~\forall~}}

\newcommand{\Sep}{\mathrm{SEP}}

\newcommand{\GHZ}{\mathrm{GHZ}}
\newcommand{\CZ}{\mathrm{CZ}}

\newcommand{\cB}{\mathcal{B}}

\newcommand{\cE}{\mathcal{E}}

\newcommand{\cG}{\mathcal{G}}
\newcommand{\cH}{\mathcal{H}}

\newcommand{\cK}{\mathcal{K}}

\newcommand{\cM}{\mathcal{M}}
\newcommand{\cN}{\mathcal{N}}
\newcommand{\cO}{\mathcal{O}}
\newcommand{\cP}{\mathcal{P}}

\newcommand{\cS}{\mathcal{S}}

\newcommand{\cY}{\mathcal{Y}}
\newcommand{\cZ}{\mathcal{Z}}

\newcommand{\dC}{\mathds{C}}

\newcommand{\bmid}{\;\big|\;}
\newcommand{\Bmid}{\;\Big|\;}

\usepackage{todonotes}

\begin{document}

\title{Statistical Methods for Quantum State Verification and Fidelity Estimation}

\author{Xiao-Dong Yu}
\email{xiao-dong.yu@uni-siegen.de}
\affiliation{Naturwissenschaftlich-Technische Fakult\"at, Universit\"at Siegen,
Walter-Flex-Str. 3, D-57068 Siegen, Germany}
\affiliation{Department of Physics, Shandong University, Jinan 250100, China}

\author{Jiangwei Shang}
\email{jiangwei.shang@bit.edu.cn}
\affiliation{Key Laboratory of Advanced Optoelectronic Quantum Architecture and 
Measurement of Ministry of Education, School of Physics, Beijing Institute of 
Technology, Beijing 100081, China}

\author{Otfried G\"uhne}
\email{otfried.guehne@uni-siegen.de}
\affiliation{Naturwissenschaftlich-Technische Fakult\"at, Universit\"at Siegen,
Walter-Flex-Str. 3, D-57068 Siegen, Germany}

\date{\today}

\begin{abstract}
  The efficient and reliable certification of quantum states is
  essential for various quantum information processing tasks as
  well as for the general progress on the implementation of quantum
  technologies. In the last few years several methods have been introduced
  which use advanced statistical methods to certify quantum states
  in a resource-efficient manner. In this article we present a review of the 
  recent
  progress in this field. We first explain how the verification and
  fidelity estimation of a quantum state can be discussed in the
  language of hypothesis testing. Then, we explain in detail 
  various strategies for the verification of entangled states
  with local measurements or measurements assisted by local operations
  and classical communication. Finally, we discuss several extensions of the
  problem, such as the certification of quantum channels and the
  verification of entanglement.
\end{abstract}

\maketitle

\section{Introduction}\label{sec:introduction}
%
A basic yet important step in quantum information processing is
the efficient and reliable characterization of quantum states. This
is not only important in certain information processing protocols,
such as quantum teleportation \cite{Bennett.etal1993}, quantum cryptography
\cite{Bennett.Brassard1984, Ekert1991, Beige.etal2002, Long.Liu2002,  
Pirandola.etal2020, Wang.etal2021}, and measurement-based quantum computation 
\cite{Raussendorf.Briegel2001,Raussendorf.etal2003}, where the state
emitted by a source needs to be characterized. The problem of state
certification arises also frequently in the technological design
and analysis of quantum devices, where the occurring quantum 
states need to be identified in an efficient manner.

Originally, a standard approach is to perform quantum state 
tomography by fully reconstructing the density matrix 
\cite{Smithey.etal1993, James.etal2001, Paris.Rehacek2004}. Tomography, 
however, is known to be both time consuming and computationally hard
due to the exponentially increasing number of parameters to be
determined \cite{Haeffner.etal2005, Shang.etal2017}. Furthermore,
in order to reconstruct a valid density matrix from experimental
data, approximations like the maximum-likelihood estimation or
Bayesian techniques have to be used \cite{Paris.Rehacek2004,
Blume-Kohout2010, Shang.etal2013}, which require additional effort 
and may lead to problematic effects \cite{Schwemmer.etal2015}.

In fact, full tomographic information is often not required,
thus a lot of effort has been devoted to characterizing quantum
states or processes with non-tomographic methods
\cite{Eisert.etal2020,Kliesch.Roth2021,Guehne.Toth2009,
Friis.etal2019,Chen.etal2021b}.  For instance, in many
experiments the fidelity of the prepared quantum state with respect to some 
target state is used as a benchmarking parameter 
\cite{Kiesel.etal2007,Kurz.etal2014,Song.etal2017}.
Consequently, various methods for the fidelity estimation and the
determination of confidence intervals have been derived
\cite{Flammia.Liu2011,Guehne.etal2007b,Seshadri.etal2021}.

\begin{figure}
  \centering
  \includegraphics[width=.45\textwidth]{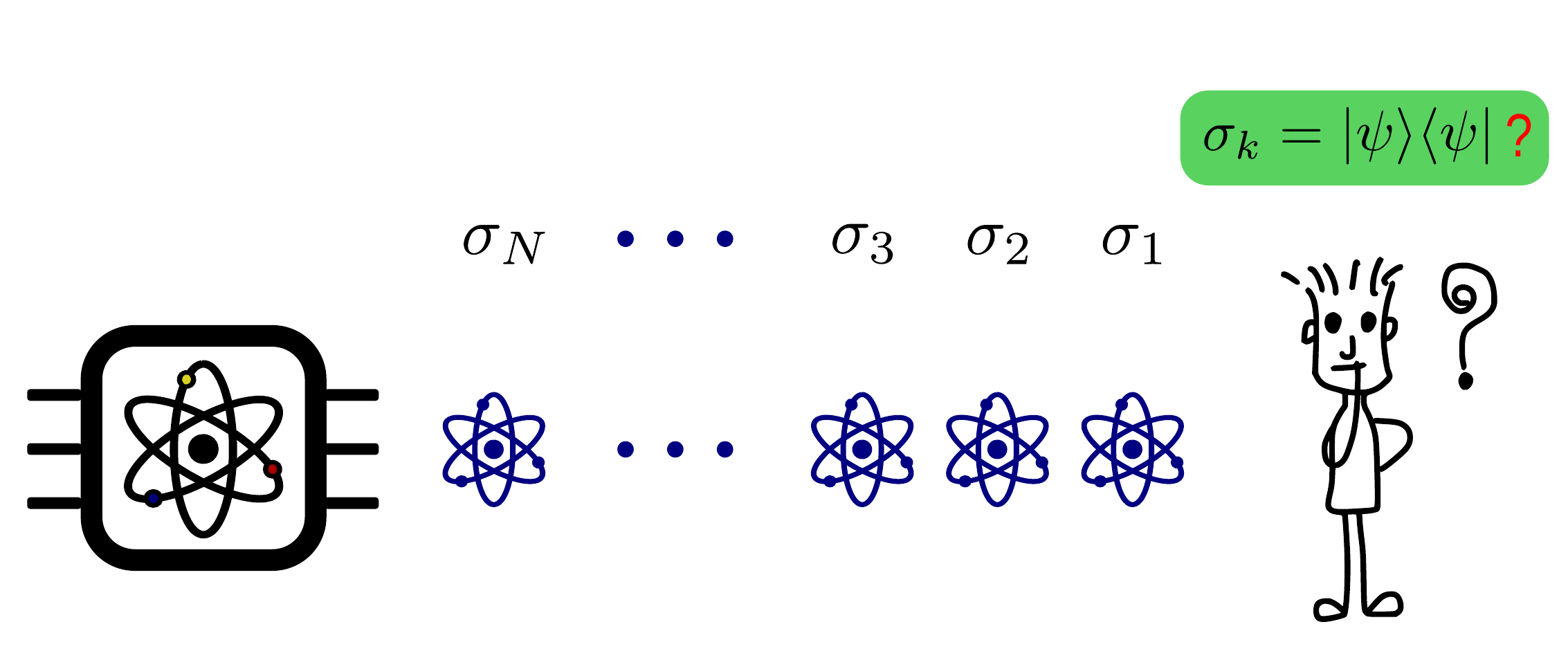}
  \caption{%
    Schematic view on quantum state verification.  One considers a quantum 
    device which is promised to produce a specific target state $\ket{\psi}$, 
    but in practice, the device produces a sequence of independent states 
    $\sigma_1,\sigma_2,\dots,\sigma_N$. The quantum state verification
    protocol studies how one can 
    verify whether $\sigma_k=\ket{\psi}\bra{\psi}$ or not in the language of 
    hypothesis testing.
    }
  \label{fig:QSV}
\end{figure}

In the last few years, the research on quantum state verification (QSV) 
has made enormous progress by using advanced statistical methods and
the framework of hypothesis testing \cite{Pallister.etal2018,Hayashi.etal2006,Hayashi2009}.
This not only leads to unambiguous statements on experimental data, but 
also results in efficient methods which require only few copies of the quantum state 
under investigation. The archetypical situation is depicted in Fig.~\ref{fig:QSV}. 
A source is promised to emit some state $\ket{\psi}.$ In practice, the device
produces a sequence of independent states $\sigma_1,\sigma_2,\dots,\sigma_N$.
How can we decide whether $\sigma_k=\ket{\psi}\bra{\psi}$ or not? What are
the optimal measurement strategies for this task, especially if not all
measurements are available due to physical constraints such as locality? 
Interestingly, for many cases these questions can be answered, and the answers 
are relevant also for experimental situations which are not as clean as 
the scenario depicted in Fig.~\ref{fig:QSV}.

In this article we review the recent developments on quantum state
verification. Our aim is to provide the reader first with a basic
and pedagogical introduction into the concepts of hypothesis testing
and state verification. Then, we explain the results for different
scenarios in details. These detailed protocols naturally depend on the
state one wishes to verify, but also on the allowed measurement protocols, 
e.g., the extent to which communication between the parties is allowed.

We are aware of the fact that statistical tools have found widespread
applications in quantum information processing and our article can
cover only a small aspect of that. There are already excellent review
articles on quantum state discrimination \cite{Barnett.Croke2009,Bae.Kwek2015}
and, very recently, on entanglement tests using witnesses from a statistical
perspective \cite{Morris.etal2021}. Furthermore, we encourage the reader
to consult the original literature on related topics, such as the estimation
of pure or mixed quantum states \cite{Bagan.etal2002, Audretsch.etal2003, 
Bagan.etal2004, Vicente.etal2010}, the estimation of drift or change point
detection \cite{Ferrie.Blume-Kohout2012, Sentis.etal2016b, Proctor.etal2020},
sequential hypothesis testing \cite{Laskowski.etal2013, Milazzo.etal2019, 
Martinez-Vargas.etal2021}, the blind channel estimation \cite{Chai.etal2020}, 
and the estimation of quantum teleportation \cite{Hu.etal2019,Cavalcanti.etal2017}.
In addition, this review focuses  on the problem of state verification in
discrete-variable quantum systems. For their counterparts, continuous-variable 
quantum systems, the verification stands as a related but independent 
problem; see Refs.~\cite{Wu.etal2021,Liu.etal2021b} for recent progresses.

This article is organized as follows. In Sec.~\ref{sec:preliminaries} we 
give an introduction to the underlying concepts. We first explain the 
framework of hypothesis testing and explain then the basic scenario of 
quantum state verification and fidelity estimation. In Sec.~\ref{sec:qsv} 
we discuss in detail the different scenarios. On the one hand, these are 
characterized by the pure bipartite or multipartite entangled state that 
should be verified. On the other hand, one can distinguish different types
of available measurements and types of communication that are allowed.
In Sec.~\ref{sec:generalizations} we discuss generalizations of
quantum state verification, such as the verification of quantum channels and
the implementation of entanglement tests based on few copies of a
state. Finally, we conclude and point out some interesting questions for
further investigation.

\section{Preliminaries and Concepts}\label{sec:preliminaries}
Before presenting the various results on QSV, we need to introduce
the required concepts. First, we discuss the notion of hypothesis testing in
some detail. Then, we can formalize the task of QSV as the main topic of this
review. Finally, we discuss the problem of estimating the fidelity with
a certain target state. Formulated as a statistical test, this task is
different from QSV, but similar from a physical point of view on aims at characterizing
the same physical quantity. Hence, methods known from fidelity estimation can
frequently be applied to QSV.

\subsection{Hypothesis testing}

Let us start by introducing the notion of hypothesis testing, which is a
method used for making statistical decisions using experimental data 
\cite{Rice2006}. Here, we focus on the Neyman-Pearson framework for
testing hypotheses \cite{Neyman.Pearson1933}. To illustrate the idea,
we consider the following classical example. Suppose that we have a coin,
which is either a fair coin with $P(\text{head})=1/2$ or a biased coin
with $P(\text{head})=3/4$. Now, we want to decide whether the coin is
fair or biased. To do this, we toss the coin $100$ times. Suppose
that we obtain more than $70$ times of heads, then a natural guess is that
the coin is biased.  The main reason for this intuition is that if the
coin were fair, then it would be very unlikely to observe the described data.
Indeed, one can directly calculate that for a fair coin the probability
of observing more than $70$ times of heads from $100$ tosses is
$p \approx 1.61 \times 10^{-5}$, while for the biased coin it is
$p \approx 0.850$, which makes this conclusion appealing.

For the general case, however, one needs a precise framework for
such conclusions, and the notion of hypothesis testing is such
a tool, giving answers on how to make a decision and how reliable
the decision is. In this framework, the above statements that the
coin is fair or biased correspond to two hypotheses. Then,
how to make the decision based on the number of heads appeared
corresponds to the decision rule. Finally, there are different types
of errors that can be made, and these errors need to be quantified. Formally, the
Neyman-Pearson framework consists of the following components:
\begin{itemize}[left=0pt]
\item
    {\bf Hypotheses.} One has two hypotheses, namely the
    \begin{itemize}
      \item
	{\it Null hypothesis $H_0$,} e.g., the hypothesis that the coin is
	fair.
      \item
	{\it Alternative hypothesis $H_1$,} e.g., the hypothesis of a specific
	unfair coin.
    \end{itemize}
  \item
  {\bf Decision rule.} This is a rule based on the observed data
  to either
    \begin{itemize}
      \item
	{\it Reject $H_0$.} In the given example, one may use the rule to reject
	$H_0$ if say, more than $60\%$ of the observed coin tosses give ``heads''.
      \item
	{\it Accept $H_0$.} Correspondingly, one accepts the assumption of 
	a fair coin if the fraction of ``heads'' is not larger than $60\%$, but 
	also other rules are conceivable.
    \end{itemize}
  \item
{\bf Errors.} For the given decision rule two errors are relevant:
    \begin{itemize}
      \item
      {\it Type I error}: This is the probability of rejecting $H_0$ when
      $H_0$ is true, that is $P(\text{type I error})=P(\text{reject $H_0$}\mid 
      H_0)$. For the example given above this error is $P_I\approx 1.76 \%$, if 
      the coin is tossed 100 times.
      \item
      {\it Type II error}: This is the probability of accepting $H_0$ when $H_1$
      is true, $P(\text{type II error})=P(\text{accept $H_0$}\mid H_1)$. For the
      example given above this is $P_{II}\approx 0.07\%$, if the coin
      is tossed 100 times.
    \end{itemize}
\end{itemize}
In this framework the hypotheses are non-symmetric: one is singled out as
the null hypothesis, denoted by $H_0$ and the other as the alternative
hypothesis, denoted by $H_1$.
Usually, $H_0$ is chosen as the hypothesis that one wants to disprove from the
experimental data. Statistically, it may happen that one makes
wrong conclusions on the accepting or rejecting, which are then characterized
by the type I/II errors. The type I error is usually called the significance 
level of the hypothesis testing. In practice, some typical values, such as 
$5\%$ or $1\%$, are widely used in various scientific fields.

Next, we explain the above notions with a discrimination task in quantum
information processing, which is similar to the previous example of coins, but
already closely related to QSV. Suppose that we have a quantum device which is
promised to always produce the same state, but it is unknown whether this state
is the basis state $\ket{0}$ or the superposition state
$\ket{+}=(\ket{0}+\ket{1})/{\sqrt{2}}$.

Now, one wishes to verify that the device is indeed producing the superposition
state $\ket{+}$ instead of the basis state $\ket{0}$. The following procedure
can be applied. First, let the null hypothesis be that the device produces the
state $\rho_0=\ket{0}\bra{0}$,  and the alternative hypothesis be that the device
produces the state $\rho_1=\ket{+}\bra{+}$. Second, one lets the device produce
$N$ copies of the state $\rho$, and performs a measurement \begin{equation}
\cM=\{\Omega,\I-\Omega\}
\end{equation}
with two outcomes on each copy.
Here, $\Omega$ is a positive semidefinite operator corresponding
to one of the two possible measurement results. In order to
discriminate the hypotheses, $\Omega$ needs to satisfy that
the probabilities of the outcomes for $\ket{0}$ and $\ket{+}$
are different, i.e., $p_0=\bra{0}\Omega\ket{0}\ne p_1:=\bra{+}
\Omega\ket{+}$.  Without loss of generality, we can assume that
$p_0 < p_1$.

Then, let $T$ denote the number of times where the result corresponding to
$\Omega$ occurs out of the $N$ measurements.  A decision rule can be defined by
\begin{equation}
\begin{cases}
    \text{reject $H_0$}\quad &\text{if}~T \ge t_0,\\
    \text{accept $H_0$}\quad &\text{if}~T < t_0.
  \end{cases}
  \label{eq:decisionRule}
\end{equation}
Here  $t_0$ is a constant that is specified \textit{before} the experiment
by the desired significance level $\alpha$, as one wishes to have
\begin{equation}
  P\bigl(T\ge t_0\bmid H_0\bigr)\le\alpha.
  \label{eq:significanceLevel}
\end{equation}
In this way, one can design a test for the described device. The main
task is then, of course, to design the measurement operator $\Omega$,
in such a way that the desired significance level can be reached
with few copies $N$.

In practice, the significance level is often not fixed from the
beginning. Instead, one characterizes the significance with the
so-called $p$-value
\begin{equation}
\delta_t:=P\bigl(T \ge t \bmid H_0 \bigr).
\label{eq:pvalue}
\end{equation}
Note that the difference between $t_0$ in Eq.~\eqref{eq:significanceLevel}
and $t$ in Eq.~\eqref{eq:pvalue} is that $t_0$ is a predefined value
that is determined before the experiment, but $t$ is the observed
result of the actual experiment. The value $1-\delta_t$ is usually called
the confidence of the hypothesis testing. If one finds a $t$ such that
$\delta_t\le\alpha$, then the null hypothesis is rejected.

Note that calculation of the $p$-value as in Eq.~\eqref{eq:pvalue} is
closely connected to the so-called large deviation bounds. In these bounds
one has a given probability distribution, here specified by the hypothesis
$H_0$, and one aims to bound the probability to find a certain deviation
from the mean value, if a statistical experiment is repeated $N$ times.
The archetypical bound of this type is the Hoeffding inequality
\cite{Hoeffding1963}, which states the following. Consider $N$ independent
(but not necessarily identically distributed) random variables
$X_i \in [a_i, b_i]$ with a mean value $\mean{X_i}.$ In a statistical
experiment, one may observe their sample mean, $X = (\sum_{i=1}^N X_i)/N.$
Then, the probability that this sample mean deviates from the overall mean
value $\mean{X} = (\sum_{i=1}^N \mean{X_i})/N$
is bounded by
\begin{equation}
  P\bigl(X-\mean{X} \geq \varepsilon\bigr) \leq \me^{-\frac{ 2 \varepsilon^2 
  N^2}{\sum_{i=1}^{N}(b_i - a_i)^2}}.
\label{eq:Hoeffding}
\end{equation}
Various similar bounds exist, such as the Bernstein, Cantelli, or McDiarmid
inequalities \cite{Bernstein1924,Cantelli1929,McDiarmid1989} and have been
frequently used to analyze quantum experiments from a statistical point of view
\cite{Flammia.Liu2011,Moroder.etal2013,Sugiyama.etal2013,Schwemmer.etal2015,Knips.etal2015, 
Sugiyama2015,Ketterer.etal2020b}.

In this article we are mainly interested in the case that the $p$-value is small
enough to reject the null hypothesis. Thus, we will not distinguish the notions
of (the probability of) the type I error, the significance level, or the
$p$-value, unless otherwise stated. The type II error defined as
\begin{equation}
  \beta:=P\bigl(T < t_0\bmid H_1\bigr),
  \label{eq:powerLevel}
\end{equation}
characterizes the power of the hypothesis testing, i.e., the smaller
the type II error is the less likely that one makes a false acceptance.

Going back to the previous example on quantum state discrimination, we may
choose $\Omega=\ket{+}\bra{+}$ and $t_0=N$, resulting in $p_0=1/2$ and $p_1=1$.
Then the decision rule will be that we accept the null hypothesis
$\rho_0=\ket{0}\bra{0}$ unless $t=N$ and the type II error of the hypothesis is
always zero. In the case that $t=N$, the null hypothesis
$\rho_0=\ket{0}\bra{0}$ is rejected in favor of the alternative hypothesis
$\rho_1=\ket{+}\bra{+}$ with confidence
\begin{equation}
1-\delta = 1-\big(\bra{0}\Omega\ket{0}\big)^N = 1-\frac{1}{2^N},
\end{equation}
where $\delta=1/2^N$ is the $p$-value.
We note that the $p$-value measures how unlikely the given data are if $H_0$ is
true. It is neither the probability that $H_0$ is false nor the probability
that $H_1$ is true.

\subsection{The basic task of quantum state verification}
With the knowledge of hypothesis testing, we can describe the model for
QSV by Pallister \textit{et al.} \cite{Pallister.etal2018}. This is one
basic model of QSV, but one should also refer to the pioneering works by Hayashi
\textit{et al.} \cite{Hayashi.etal2006,Hayashi2009} using slightly different assumptions.

Suppose that we have a quantum device which is promised to produce a
specific target state $\ket{\psi}$, for instance the entangled singlet
state $\ket{\psi^-}= (\ket{01} -\ket{10})/\sqrt{2}$. In practice, the
device produces a sequence of independent states
$\sigma_1,\sigma_2,\dots,\sigma_N$. The QSV protocol studies how one can verify
whether $\sigma_k=\ket{\psi}\bra{\psi}$ in the language of hypothesis testing;
see Fig.~\ref{fig:QSV}. To this end, we choose the alternative hypothesis as
\begin{equation}
  H_1:~\sigma_k\in S:=\{\ket{\psi}\bra{\psi}\}~\text{for all}~k.
  \label{eq:QSValt}
\end{equation}
A first simple, but naive, choice of the null hypothesis could be
$\sigma_k\in S^c= \{\rho\mid\rho\ne\ket{\psi}\bra{\psi}\}$, where
$S^c$ denotes the complement of $S$. This, however, is not a good choice,
because for any $\rho \in S$ there exists a $\tilde \rho\in S^c$ such that
$\tilde \rho$ can be arbitrarily close to $\rho$. As a result, it is impossible to
reject the null hypothesis with a strictly positive confidence.

Thus, instead of choosing $S^c$ as the null hypothesis, we choose
the set of states that is $\varepsilon$-away from the target state
$\ket{\psi}$, i.e.,
\begin{equation}
H_0:~\sigma_k\in S_\varepsilon:=\bigl\{\rho\bmid\bra{\psi}\rho\ket{\psi}\le
  1-\varepsilon\bigr\}~\text{for all}~k.
  \label{eq:QSVnull}
\end{equation}
If this hypothesis $H_0$ is rejected, this implies that at least some of the
states $\sigma_k$ are close to the target state $\ket{\psi}$.

The basic QSV protocol from Ref.~\cite{Pallister.etal2018} considers
the idealized scenario where either $H_0$ is true or $H_1$ is true.
This may sound unrealistic, but this idealized model is more convenient
for theoretical studies. Moreover, the results can be directly
generalized to the practically relevant estimation of fidelities
described in the next subsection.  We also note that in the QSV
protocol, the states $\sigma_k$ generated by the device are only
assumed to be independent, but not necessarily to be identical, i.e.,
arbitrary fluctuation of $\sigma_k$ is allowed as long as
$\sigma_{1,2,\dots,N}=\sigma_1\otimes\sigma_2\otimes\dots\otimes\sigma_N$.
This is relevant, as in realistic scenarios there may be some
drift in the source, leading to a systematic change of the states
$\sigma_k$ with time. The generalization of the QSV protocols to
non-independent sources, the so-called adversarial scenario, will be discussed
in Sec.~\ref{sec:generalizations}.

Due to the trade-off between type I and type II errors, different
figures of merit for a hypothesis testing can be chosen depending on
different physical or mathematical considerations.  In QSV,
the type II error is usually constrained to be zero, which
means if the quantum device indeed produces the desired state
$\ket{\psi}$, then it will always pass the test. This assumption
is, however, not necessary. We will talk about possible
generalizations in Sec.~\ref{sec:generalizations}.

Then, it finally remains to discuss the measurements that
shall be performed. In general, for each state $\sigma_k$,
the verifier may apply a measurement, $\{\Omega_\ell,\I-\Omega_\ell\}$
randomly chosen from some set with probability $p_\ell$. For instance, if the
singlet state $\ket{\psi^-}$ shall be verified, he/she may perform spin
measurements in correlated, but arbitrary directions,  in order to observe the
anti-correlations which are characteristic of the singlet state.

Thus, any QSV strategy can be expressed as an overall measurement
\begin{equation}
\Omega=\sum_{\ell=1}^m p_\ell\Omega_\ell,
\label{eq:verificationStragety}
\end{equation}
where $(p_1,p_2,\dots,p_m)$ is a probability distribution, and
$\{\Omega_\ell,\I-\Omega_\ell\}$ are allowed measurements with
outcomes labeled by ``pass'' and ``fail'', respectively.
An essential insight in the following discussion is that often
only the properties of $\Omega$ are
relevant, but not the specific forms of $\{p_\ell,\Omega_\ell\}$.

To guarantee that the type II error is zero, i.e., that the target
state $\ket{\psi}$ never fails the test, all $\Omega_\ell$ are
required to satisfy that
\begin{equation}
 \bra{\psi}\Omega_\ell\ket{\psi}=1
 ~\Leftrightarrow~
 \Omega_\ell\ket{\psi}=\ket{\psi},
 \label{eq:OmegaEigen}
\end{equation}
where the equivalence follows from the fact that the measurement
effect $\Omega_\ell$ cannot have eigenvalues larger than one. In
a pass instance, the verifier continues to state $\sigma_{k+1}$
and repeats the test, otherwise the verification ends and the verifier accepts
the hypothesis $H_0$, i.e., he/she asserts that the states were not
$\ket{\psi}$.  If all the $N$ states $\sigma_k$ pass the test, then the
verifier rejects the hypothesis $H_0$ in favor of $H_1$, asserts that the
states were indeed $\ket{\psi}$
and that the quantum device is working as intended.

To evaluate the type I error of the hypothesis testing scheme,
we consider the worst-case failure probability
$\max_{\bra{\psi}\sigma\ket{\psi}\le 1-\varepsilon}\Tr(\Omega\sigma)$
of each run. Note that $\Omega\ket{\psi}=\ket{\psi}$ and the maximal
eigenvalue of $\Omega$ equals one. So, one can restrict $\sigma$ to be of the following 
form for the maximization
\begin{equation}
  \sigma=(1-\varepsilon')\ket{\psi}\bra{\psi}
  +\varepsilon'\ket{\psi^\perp}\bra{\psi^\perp},
\end{equation}
where $\varepsilon'\ge\varepsilon$ and $\ket{\psi^\perp}$ is orthogonal to
$\ket{\psi}$. This further implies that the maximization is achieved when
$\varepsilon'=\varepsilon$ and $\ket{\psi^\perp}$ is the eigenvector corresponding
to the second largest eigenvalue of $\Omega$. Thus, the worst-case failure
probability is given by
\begin{equation}
  \max_{\bra{\psi}\sigma\ket{\psi}\le 1-\varepsilon}\Tr(\Omega\sigma)
  =1-\varepsilon \nu(\Omega),
  \label{eq:worstcase}
\end{equation}
where $\nu(\Omega)$ represents the spectral gap between the largest
($\lambda_1=1$) and the second largest ($\lambda_2 =1-\nu$)
eigenvalues of $\Omega$. In the case that $H_0$ is true and all
the $N$ sample states still pass the test, the type I error is
bounded by
\begin{equation}
  \delta\le[1-\varepsilon \nu(\Omega)]^N,
  \label{eq:failureProbability}
\end{equation}
Thus, to achieve a given confidence $1-\delta$, it is sufficient to
take
\begin{equation}
N=\frac{\ln(\delta)}{\ln[1-\varepsilon\nu(\Omega)]}
\label{eq:sampling}
\end{equation}
sample states from the quantum device. In the high precision limit
($\varepsilon,\delta\to 0$) this scales as
\begin{equation}
  N\approx[\nu(\Omega)]^{-1}\varepsilon^{-1}\ln(\delta^{-1}).
  \label{eq:samplingComplexity}
\end{equation}

For the detailed construction of state verification protocols for
specific states the main problem is to find the optimal $\Omega$.
Here, the optimization is typically subject to some constraints, as
not all measurements are available. This will be discussed in details
in the remainder of this review.

\subsection{Quantum fidelity estimation}
As we have mentioned, the QSV scenario above is an idealized model.
This is because the choice of the two hypotheses in
Eqs.~(\ref{eq:QSValt},\,\ref{eq:QSVnull}) is impractical, in fact the
hypothesis that $\sigma_k=\ket{\psi}\bra{\psi}$ for all $k$ is very unlikely to
be true due to the unavoidable noise in actual experiments. Hence, a more
practical model is to characterize the average fidelity of the output of the
quantum device, which we refer to as quantum fidelity estimation (QFE)
\cite{Yu.etal2019};
see also a related statistical entanglement witness method in
Refs.~\cite{Dimic.Dakic2018,Saggio.etal2019} or Sec.~\ref{sec:generalizations}.

To explain this, recall that the fidelity of some mixed quantum
state $\rho$ with a pure target state $\ket{\psi}$ is given by
\begin{equation}
  F_\psi(\rho) = \bra{\psi} \rho \ket{\psi} = \Tr(\rho \ketbra{\psi}).
\end{equation}
If $\rho$ is a state generated in an experiment, this fidelity is a
frequently used
parameter to compare different implementations 
\cite{Kiesel.etal2007,Kurz.etal2014,Song.etal2017}. Also, the fidelity
may be used to prove that the state
$\rho$ is entangled: If the fidelity exceeds the maximal value
for separable states, then the state $\rho$ must necessarily
be entangled. This approach is widely used \cite{Guehne.Toth2009}, although not
all forms of entanglement can be detected with it
\cite{Weilenmann.etal2020,Guehne.etal2021}.

One of the main tasks for experiments is to develop methods to
determine or estimate the fidelity from few measurements,
without doing full state tomography of the state $\rho$. Indeed,
for many cases, such as Bell states, cluster and graph states, or Dicke states,
methods are known to be able to achieve this efficiently, i.e., without using 
an exponentially increasing number of measurements
\cite{Toth.Guehne2005,Toth.Guehne2005b, Toth2007, Guehne.etal2007b, Toth.etal2010}.
Also known are the statistical tests based on the Hoeffding inequality of 
Eq.~\eqref{eq:Hoeffding} \cite{Flammia.Liu2011,Schwemmer.etal2015}, which allow 
to compute rigorous error bars for the case that only a finite number of copies 
of the state $\rho$ are available.

How can these concepts be connected with the concept of QSV?
Given the output states $\sigma_k$, one can take the null and
alternative hypotheses as follows:
\begin{align}
  \label{eq:fidelityWitnessNull}
  H_0:&~
  \frac{1}{N}\sum_{k=1}^N\bra{\psi}\sigma_k\ket{\psi}\le 1-\varepsilon,\\
  \label{eq:fidelityWitnessAlt}
  H_1:&~
  \frac{1}{N}\sum_{k=1}^N\bra{\psi}\sigma_k\ket{\psi}> 1-\varepsilon.
\end{align}
These hypotheses are essentially statements about the average fidelity
of the states $\sigma_k$ with the target state $\ket{\psi}$. In order to
test these hypotheses, the verifier can take the same measurement strategy
as in Eq.~\eqref{eq:verificationStragety}. In this case, the verifier
performs a random measurement $\{\Omega_\ell,\I-\Omega_\ell\}$ for each
state $\sigma_k$ and calculates the frequency $f$ of the pass instances,
i.e.,
\begin{equation}
  f=\frac{t}{N},
  \label{eq:frequency}
\end{equation}
where $t$ is the number of cases where $\{\sigma_k\}_{k=1}^N$ pass the
test. Now, the  verifier may reject the null hypothesis and conclude that
the average fidelity is larger than $1-\varepsilon$ if
\begin{equation}
f>1-\varepsilon\nu(\Omega). \label{eq:fidelityDecision}
\end{equation}
Furthermore, given the independence of $\sigma_k$, the confidence $1-\delta$
can be determined by the Chernoff-Hoeffding theorem \cite{Hoeffding1963}
\begin{equation}
  \delta\le\me^{-D[f\|1-\varepsilon\nu(\Omega)]N},
  \label{eq:CHbound}
\end{equation}
where
\begin{equation}
  D(x\|y)=x\ln\biggl(\frac{x}{y}\biggr)+(1-x)\ln\biggl(\frac{1-x}{1-y}\biggr)
\end{equation}
is the Kullback–Leibler divergence. Note that the Hoeffding bound in
Eq.~\eqref{eq:Hoeffding} can also be used for deriving the confidence, but it
is weaker than Eq.~\eqref{eq:CHbound}.
Especially, if all the $N$ states pass the test, i.e., $f=1$,
Eq.~\eqref{eq:CHbound} reduces to Eq.~\eqref{eq:failureProbability}, but now
the physical consequences
are clear: This is the confidence that the \textit{average} fidelity of the
output of the quantum device is larger than $1-\varepsilon$. Before, the
rejection of the null hypothesis was interpreted as a statement on some
of the states $\sigma_k$ (see the discussion after Eq.~(\ref{eq:QSVnull}))
or, if one of the hypotheses in Eqs.~(\ref{eq:QSValt},\,\ref{eq:QSVnull})
is assumed to be true, as an acceptance of the hypothesis
that $\sigma_k=\ket{\psi}\bra{\psi}$ for all $k$.

A key feature of the presented QSV and QFE protocols is that the failure 
probability $\delta$ decreases exponentially with $N$, hence the target
state $\ket{\psi}$ can be potentially verified using only few copies of the 
state. As seen from Eqs.~(\ref{eq:failureProbability},\,\ref{eq:CHbound}),
the performance of a verification strategy depends solely on $\nu(\Omega)$.  
Therefore, to achieve an optimal strategy, we need to maximize $\nu(\Omega)$
over all accessible measurements.

The previous discussions also imply that all measurement settings for QSV can
be directly used for QFE. For simplicity, hereafter, we will only consider QSV
strategies, but these strategies can be directly used for QFE, unless otherwise
stated. Note, however, that optimality statements for an QSV strategy do not
necessarily imply optimality for the QFE problem. This is because the bound in 
Eq.~\eqref{eq:CHbound} may not be optimal when $f<1$.

\section{QSV for entangled states}\label{sec:qsv}

In QSV, a fundamental quantity is the spectral gap $\nu(\Omega)$. For
any quantum state $\ket{\psi}$, if there is no constraint to the choice of
measurements, one can easily see that the optimal verification strategy is to
take $\Omega=\ket{\psi}\bra{\psi}$, and correspondingly, $\nu(\Omega)=1$. If
the quantum system involves many parties, it is, however, difficult or even
impossible to implement the entangled measurement
$\{\ket{\psi}\bra{\psi},\I-\ket{\psi}\bra{\psi}\}$. In this case, a more
realistic model is to consider local measurements or measurements assisted
by local operations and classical communication (LOCC).

By a strategy with local measurements we mean that different parties
perform their measurements independently, that is, no communication is
needed during the measurements. For a strategy with measurements assisted
by LOCC, some parties perform the measurements first, then the measurement 
outcomes are sent to the other parties, whose measurements depend on the
received outcomes. As a result, QSV with local measurements and LOCC 
measurements are also called the nonadaptive and adaptive scenarios, 
respectively. Note that the adaptivity here does not mean that the verification 
strategy $\Omega$ in the $k$-th run is changed based on the results of the
previous $k-1$ tests.

\subsection{QSV with local measurements}

\begin{figure}
  \centering
  \includegraphics[width=.45\textwidth]{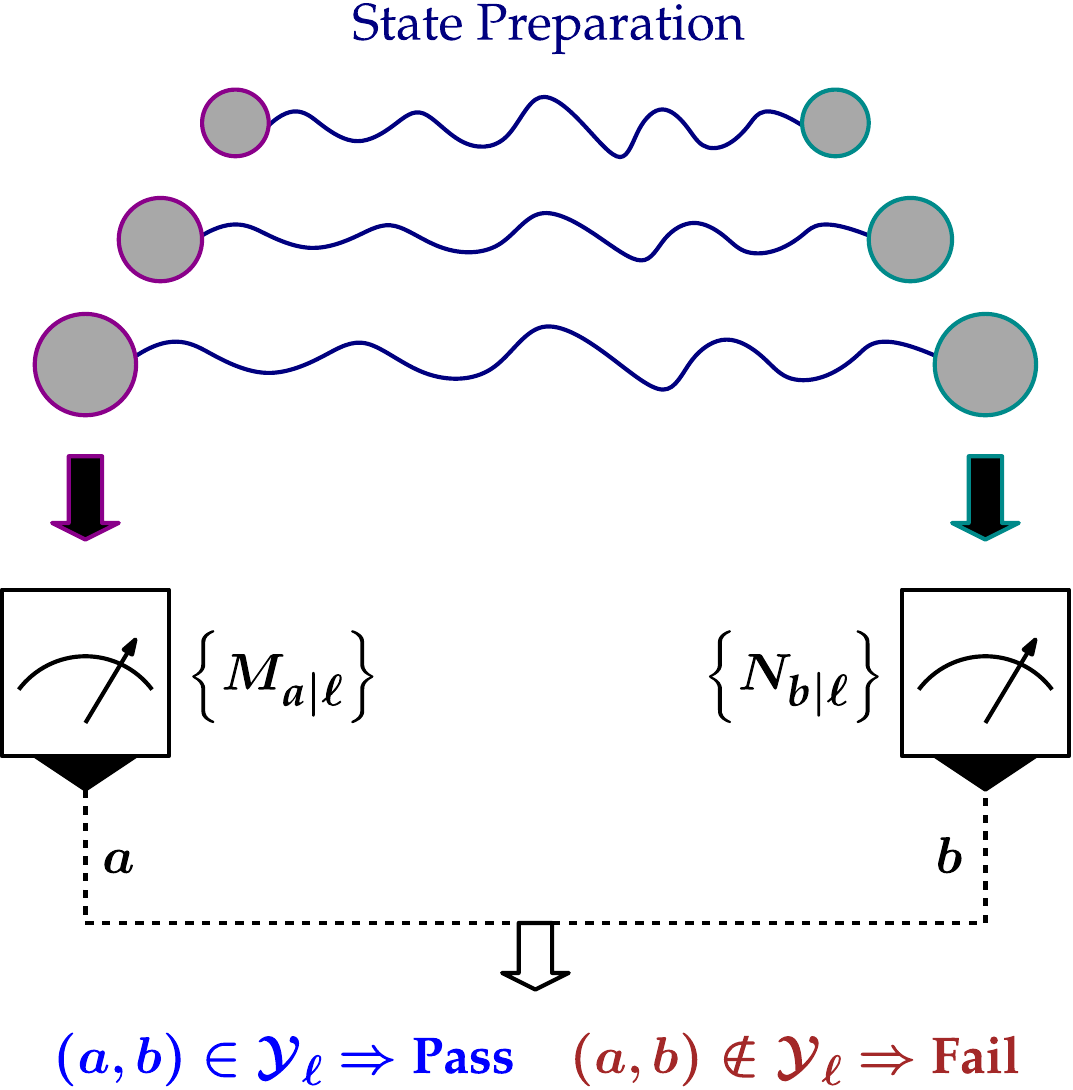}
  \caption{%
    QSV with local measurements. In the $k$-th run, Alice and Bob
    perform some random measurements $\cM_\ell=\{M_{a|\ell}\}_{a=1}^{d_A}$ and
    $\cN_\ell=\{N_{b|\ell}\}_{b=1}^{d_B}$ with probability $p_\ell$ using 
    shared randomness. The state $\sigma_k$ passes the test if their 
    measurement outcomes $a,b$ satisfy that $(a,b)\in\cY_\ell$. Note that 
    classical communication is still necessary for making the decision on pass 
    or fail.
  }
  \label{fig:local}
\end{figure}

In Ref.~\cite{Pallister.etal2018}, the verification of entangled pure
states with local projective measurements was introduced, as illustrated
in Fig.~\ref{fig:local}. The quantum device prepares quantum states $\sigma_k$
for $k=1,2,\dots,N$, which are promised to be an entangled state $\ket{\psi}$
between Alice and Bob. Suppose that each of the parties performs a single local
projective measurements, e.g., $\cM=\{M_1,M_2,\dots,M_{d_A}\}$ and
$\cN=\{N_1,N_2,\dots,N_{d_B}\}$, then the entangled state $\ket{\psi}$ cannot be
verified.  This is because $M_a\otimes N_b$ are orthogonal projectors, and thus
there exists no projector of the form
\begin{equation}
\Omega=\sum_{(a,b)\in\cY}M_a\otimes N_b
\end{equation}
that can single out $\ket{\psi}$ as a nondegenerate eigenvector,
where $\cY$ is any subset of the measurement outcomes.
However, if we
consider randomly chosen measurements and construct different projectors
$\Omega_\ell$, then it is possible to get a nonzero spectral gap $\nu(\Omega)$
for
\begin{equation}
  \Omega=\sum_\ell p_\ell\Omega_\ell.
  \label{eq:localStrategy}
\end{equation}
Indeed, the previously mentioned example of spin correlation
measurements on the singlet state illustrates already the potential
advantage of randomly chosen measurements.

More formally, let Alice and Bob perform local measurements
$\cM_\ell=\{M_{1|\ell},M_{2|\ell},\dots,M_{{d_A}|\ell}\}$ and
$\cN_\ell=\{N_{1|\ell},N_{2|\ell},\dots,N_{{d_B}|\ell}\}$ with some probability
$p_\ell$, where both $M_{a|\ell}$
and $N_{b|\ell}$ add up to the identity.  Let the set
$\cY_\ell$ be defined as
\begin{equation}
  \cY_\ell=\Bigl\{(a,b)\Bmid M_{a|\ell}\otimes
    N_{b|\ell}\ket{\psi}\ne 0\Bigr\},
  \label{eq:cY}
\end{equation}
then $\Omega_\ell$ may be chosen as
\begin{equation}
  \Omega_\ell=\sum_{(a,b)\in\cY_\ell}M_{a|\ell}\otimes N_{b|\ell}.
  \label{eq:OmegaLocal}
\end{equation}
From the definition of $\cY_\ell$, it follows directly that
$\Omega_\ell\ket{\psi}=\ket{\psi}$ and thus the verification
defined as in Eq.~\eqref{eq:localStrategy} also satisfies that
$\Omega\ket{\psi}=\ket{\psi}$.  In order to make $\Omega_\ell$
nontrivial, i.e., $\Omega_\ell\ne\I$, there must
exist some $M_{a|\ell}$ and $N_{b|\ell}$ such
that $M_{a|\ell}\otimes N_{b|\ell}\ket{\psi}=0$.

Thus, finding the optimal local strategy can be written as
\begin{equation}
  \begin{aligned}
    &\maxover[p_\ell,\,\Omega_\ell] \quad &&\nu(\Omega)\\
    &\subto &&\Omega=\sum_{\ell=1}^m p_\ell\Omega_\ell,\\
    &       &&\Omega_\ell=\sum_{(a,b)\in\cY_\ell}M_{a|\ell}\otimes
    N_{b|\ell}~~\FA \ell,\\
    &       &&\sum_{\ell=1}^m p_\ell=1,~~p_\ell\ge 0~\FA\ell.
  \end{aligned}
  \label{eq:optimizationLocal}
\end{equation}
Constructing the optimal verification strategy with local measurements is in
general difficult, as the set of local strategies is complicated. However,
efficient/optimal local strategies have been constructed for various
widely-used states in quantum information processing, and in the following we
will discuss these in details.

\subsubsection{Bell states}
As the first example, we consider the verification of Bell states
\begin{equation}
  \ket{\psi}=\frac{1}{\sqrt{2}}\Big(\ket{00}+\ket{11}\Big).
  \label{eq:Bell}
\end{equation}
In order to construct the local projectors, we take advantage of the fact that
Bell states are eigenstates of local observables,
\begin{equation}
  X\otimes X\ket{\psi}=\ket{\psi},~
  Y\otimes Y\ket{\psi}=-\ket{\psi},~
  Z\otimes Z\ket{\psi}=\ket{\psi},
  \label{eq:BellStab}
\end{equation}
where $X,Y,Z$ are the Pauli matrices. The fact that certain
pure quantum states are uniquely defined as eigenstates of such observables is
described more generally in the so-called stabilizer formalism
\cite{Nielsen.Chuang2000}; see also below.

With this relation, we can construct the verification strategy 
\cite{Pallister.etal2018}
\begin{equation}
  \Omega=\frac{1}{3}P_{XX}^++\frac{1}{3}P_{YY}^-+\frac{1}{3}P_{ZZ}^+,
  \label{eq:BellStrategy}
\end{equation}
where
\begin{equation}
  P_{XX}^+=P_X^+\otimes P_X^++P_X^-\otimes P_X^-
  =\ket{+}\bra{+}^{\otimes 2}+\ket{-}\bra{-}^{\otimes 2},
\end{equation}
and similar for $P_{YY}^-$ and $P_{ZZ}^+$.

In experiments, the verification protocol works as follows:
In each run, Alice and Bob first use shared randomness to
select with equal probability which measurement, $X\otimes X$,
$Y\otimes Y$, or $Z\otimes Z$, they wish to perform. Second,
they perform the corresponding measurements independently and
share the measurement outcomes with classical communication.
The result of this run is then decided by comparing their
outcomes. For the $Y\otimes Y$ ($X\otimes X$ or $Z\otimes Z$)
measurement, if their outcomes are different (the same), the
outcome is labeled as pass, otherwise labeled as fail. At last,
the frequency of the pass instances can be calculated after all
the $N$ runs.

From Eq.~\eqref{eq:BellStrategy}, one can easily verify that
the spectral gap is
\begin{equation}
  \nu(\Omega)=\frac{2}{3}.
  \label{eq:BellSpec}
\end{equation}
As proved in Ref.~\cite{Pallister.etal2018}, this is the largest
spectral gap achievable with projective local measurements. Moreover,
as proved in Refs.~\cite{Wang.Hayashi2019,Yu.etal2019}, this is also
optimal even if LOCC measurements are considered.

\subsubsection{Maximally entangled two-party states}%
The verification of Bell states can be directly generalized
to the two-qudit maximally entangled states. For the two-qudit
maximally entangled state
\begin{equation}
  \ket{\psi}=\frac{1}{\sqrt{d}}\sum_{\alpha=1}^{d}\ket{\alpha\alpha},
  \label{eq:MES}
\end{equation}
an important symmetry \cite{Hayashi.etal2006,Hayashi2009} that can be used for
constructing the verification strategy is that
\begin{equation}
  U\otimes U^*\ket{\psi}=\ket{\psi},
  \label{eq:MESsymmetry}
\end{equation}
where $U\in SU(d)$ and $U^*$ is the complex conjugate of
$U$ in the computational basis $\{\ket{\alpha}\}_{\alpha=1}^{d}$.
Let
\begin{equation}
  P_{ZZ}=\sum_{\alpha=1}^{d}\ket{\alpha}\bra{\alpha}\otimes
  \ket{\alpha}\bra{\alpha},
  \label{eq:Pdelta}
\end{equation}
then $P_{ZZ} \ket{\psi} = \ket{\psi}$ and a verification strategy
with a continuous family of measurements can be written as
\begin{equation}
  \Omega=\int_U U\otimes U^* P_{ZZ} U^\dagger\otimes U^T\md U
  =\frac{\I\otimes\I+d\ket{\psi}\bra{\psi}}{d+1},
  \label{eq:MESStrategyCont}
\end{equation}
where the integral is with respect to the Haar measure. An easy way
to see the second equality in Eq.~\eqref{eq:MESStrategyCont} is that
the partial transpose of $\Omega$ is a Werner state \cite{Werner1989}.
The spectral gap of this strategy is given by
\begin{equation}
  \nu(\Omega)=\frac{d}{d+1}.
  \label{eq:MESSpec}
\end{equation}
As proved in Ref.~\cite{Zhu.Hayashi2019}, this strategy is optimal not only
over all local measurements, but also over all LOCC measurements.

Equation~\eqref{eq:MESStrategyCont} needs a continuous family  of
measurements, but it also admits a representation with finite
number of measurements. This can be seen as follows. As $\Omega$
belongs to the convex hull of $\{U\otimes U^* P_{ZZ} U^\dagger\otimes
U^T\}_U$ and the underlying space is finite-dimensional, the well-known
Minkowski-Carath\'eodory theorem \cite{Simon2011} implies that there
exists a finite set $\{U_\ell\}_{\ell=1}^m$ such that
\begin{equation}
  \Omega=\sum_{\ell=1}^m p_\ell U_\ell\otimes U_\ell^* P_{ZZ}
  U_\ell^\dagger\otimes U_\ell^T,
  \label{eq:MESStrategyDisc}
\end{equation}
for some probability distribution $(p_1,p_2,\dots,p_m)$.
An explicit construction of Eq.~\eqref{eq:MESStrategyDisc} based on mutually
unbiased bases \cite{Durt.etal2010} or weighted complex projective $2$-designs
\cite{Zauner2011,Roy.Scott2007} can be found in Ref.~\cite{Zhu.Hayashi2019}. An
alternative construction based on measurements in the computational and
Fourier bases accompanied with random local phase gates is presented in the next
subsection.

\subsubsection{GHZ states}%
The QSV protocol is also applicable to multi-party entangled states. We take
the Greenberger-Horne-Zeilinger (GHZ) states, which are among the most widely
used states in quantum information processing, as examples to illustrate this.
The $n$-qubit GHZ state is defined as
\begin{equation}
  \ket{\GHZ_n}=\frac{1}{\sqrt{2}}
  \Big(\ket{0}^{\otimes n}+\ket{1}^{\otimes n}\Big).
  \label{eq:GHZ}
\end{equation}
It is well-known that the $n$-qubit GHZ state is identified by
\begin{align}
  \label{eq:GHZgen1}
  &X\otimes X\otimes\dots\otimes X\ket{\GHZ_n}=\ket{\GHZ_n},\\
  \label{eq:GHZgen2}
  &Z\otimes\I^{\otimes k}\otimes
  Z\otimes\I^{\otimes(n-k-2)}\ket{\GHZ_n}=\ket{\GHZ_n},
\end{align}
for $k=0,1,2,\dots,n-2$. In the language of the stabilizer formalism
\cite{Nielsen.Chuang2000}, the $n$ operators involved in
Eqs.~(\ref{eq:GHZgen1},\,\ref{eq:GHZgen2}) are called generators of the
stabilizer group, which consists of the elements (besides arbitrary
permutations)
\begin{align}
  \label{eq:GHZstab1}
  &Z^{\otimes 2k}\otimes\I^{\otimes(n-2k)},\\
  \label{eq:GHZstab2}
  &(-1)^kY^{\otimes 2k}\otimes X^{\otimes(n-2k)},
\end{align}
for $k=0,1,2,\dots,\lfloor n/2\rfloor$. The group is called the stabilizer
group for $\ket{\GHZ_n}$ because all its $2^n$ elements $g$ satisfy that
$g\ket{\GHZ_n}=\ket{\GHZ_n}$. The elements of the stabilizer group play
an outstanding role in the construction of quantum error-correcting code 
\cite{Gottesman1997}, Bell inequalities
\cite{Mermin1990b,Guehne.etal2005}, and entanglement detection
\cite{Toth.Guehne2005b}.

For the verification of GHZ states different problems arise.
Besides asking for the optimal verification strategy one may ask for
efficient strategies, in the sense that these do not require many
measurements. Concerning the latter point and  in analogy to the results
on entanglement witnesses \cite{Toth.Guehne2005}, Zhu and Hayashi
showed that any $n$-qubit GHZ stated can be verified with two
measurement settings~\cite{Zhu.Hayashi2018},
\begin{equation}
  X\otimes X\otimes \cdots \otimes X,~~~
  Z\otimes Z\otimes \cdots \otimes Z.
  \label{eq:GHZ2}
\end{equation}
The key observation is that if the prepared state is indeed the GHZ state
$\ket{\GHZ_n}$ and the $n$ parties perform the measurement $Z^{\otimes n}$, then
either all the $n$ parties get the outcome $+1$ or all of them get the outcome
$-1$. This will actually force the possible states to be within the subspace
spanned by $\{\ket{0}^{\otimes n},\ket{1}^{\otimes n}\}$. Then, the measurement
$X^{\otimes n}$ can distinguish the two states $(\ket{0}^{\otimes
n}+\ket{1}^{\otimes n})/\sqrt{2}$ and $(\ket{0}^{\otimes n}-\ket{1}^{\otimes
n})/\sqrt{2}$. Correspondingly, the verification operator can be written as
\begin{equation}
  \Omega=\frac{1}{2}P_{Z^{\otimes n}}
  +\frac{1}{2}P_{X^{\otimes n}}^+,
  \label{eq:GHZ2Strategy}
\end{equation}
where
\begin{align}
  \label{eq:Pzn}
  &P_{Z^{\otimes n}}
  =\ket{0}\bra{0}^{\otimes n}+\ket{1}\bra{1}^{\otimes n},\\
  &P_{X^{\otimes n}}^+=\frac{\I+X^{\otimes n}}{2}.
\end{align}
Physically, $P_{X^{\otimes n}}^+$ means that the test is passed when the product of
the measurement outcomes of the $n$ different parties is one. The spectral gap is
\begin{equation}
  \nu(\Omega)=\frac{1}{2}.
  \label{eq:GHZ2Spec}
\end{equation}

In the previous verification strategy, only the Pauli $X$ and $Z$
measurements are involved, but it is not optimal. In Ref.~\cite{Li.etal2020},
Li \textit{et al.} showed that the optimal strategy can be achieved by 
considering all the stabilizers in Eq.~\eqref{eq:GHZstab2}.  By choosing the 
measurement settings (besides the permutations)
\begin{equation}
  Z^{\otimes n},\quad
  Y^{\otimes 2k}\otimes X^{\otimes n-2k},~~k=0,1,\dots,\lfloor n/2\rfloor.
  \label{eq:GHZOptimalSetting}
\end{equation}
The optimal strategy can be achieved by
\begin{equation}
  \Omega=\frac{1}{3}\biggl(P_{Z^{\otimes n}}
  +\frac{1}{2^{n-2}}\sum_{g\in\cS_{XY}}P_g^+\biggr),
  \label{eq:GHZOptimalStrategy}
\end{equation}
where $P_{Z^{\otimes n}}$ is defined by Eq.~\eqref{eq:Pzn}, $\cS_{XY}$ is the
set of all the $2^{n-1}$ stabilizers of the form in Eq.~\eqref{eq:GHZstab2},
and
\begin{equation}
  P_g^+=\frac{\I+g}{2}
  \label{eq:stabProj}
\end{equation}
means that the product of the measurement outcomes of the $n$ different parties
is one. The spectral gap
\begin{equation}
  \nu(\Omega)=\frac{2}{3}
  \label{eq:GHZOptimalSpec}
\end{equation}
is optimal not only over all local measurements but also over all LOCC
measurements. The above strategy can also be generalized to the $n$-qudit GHZ
state with the optimal spectral gap $\nu(\Omega)=d/(d+1)$ \cite{Li.etal2020}.

At last, we would like to mention that it is also possible to verify GHZ states
in the presence of dishonest parties. See a pioneering work by Pappa
\textit{et al.} \cite{Pappa.etal2012}, a recent result by Han \textit{et al.} 
\cite{Han.etal2021}, and also a completely device-independent
treatment by Go\v{c}anin \textit{et al.} \cite{Gocanin.etal2022}.

\subsubsection{Graph and hypergraph states}
The verification of GHZ states can be generalized to a class of widely-used
entangled states in quantum information, the so-called stabilizer states
\cite{Gottesman1997}.  Without loss of generality, we only need to consider
the family of graph states \cite{Duer.Briegel2004,Hein.etal2004}, as any
stabilizer state is equivalent to a graph state up to a local Clifford (LC)
operation \cite{VandenNest.etal2004,Grassl.etal2002,Schlingemann2002}. Here,
local Clifford operations are local unitary transformations, which leave the
set of Pauli matrices up to some signs invariant.

\begin{figure}[t]
  \centering
  \includegraphics[width=.45\textwidth]{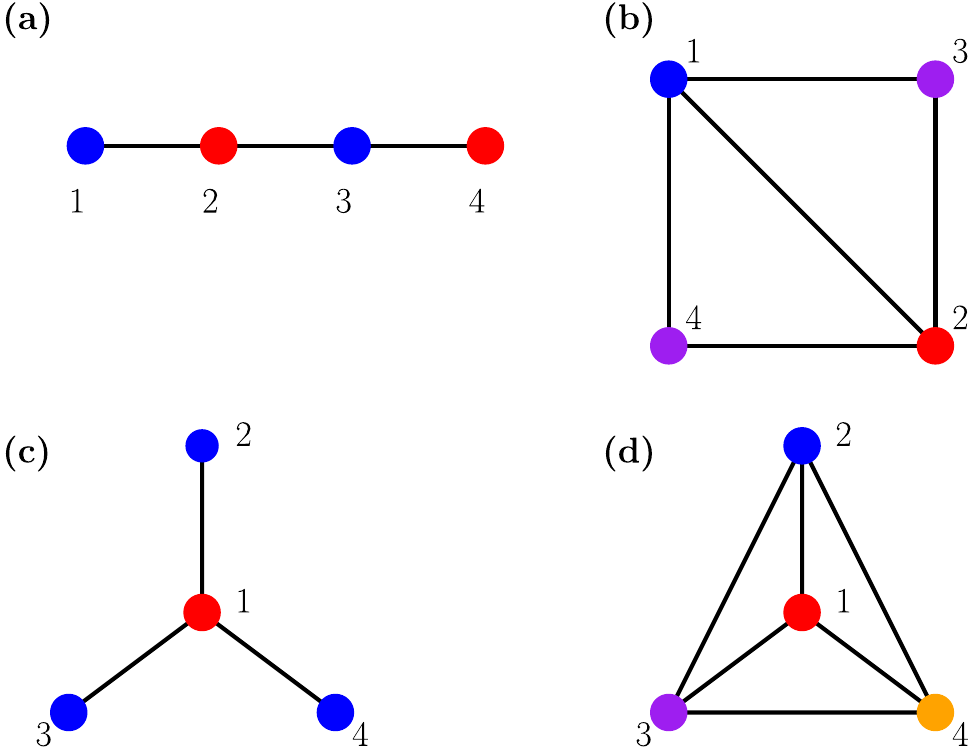}
  \caption{%
    Some graphs and their optimal coloring. Graphs (a) and (b) have equivalent
    corresponding graph states (up to an LC operation), but their chromatic
    numbers are different. So are graphs (c) and (d), and moreover they are
    equivalent to the $4$-qubit GHZ state.
  }
  \label{fig:graph}
\end{figure}

Graph states are defined as follows: First, one considers a graph $G=(V,E)$, 
i.e., an object with $n:=\abs{V}$ vertices and $\abs{E}$ edges connecting some 
pairs of the vertices; see some examples in Fig.~\ref{fig:graph}. We can 
associate to each vertex $i\in V$ an operator
\begin{equation}
  g_i=X_i\otimes\bigotimes_{\{i,j\}\in E}Z_j,
  \label{eq:GraphStab}
\end{equation}
that is, $g_i$ consists of a Pauli $X$ observable on the qubit $i$ and
a Pauli $Z$ observable on all qubits in the neighbourhood. The graph
state $\ket{G}$ corresponding to the graph $G$ is then defined as the
unique state that satisfies the eigenvalue relations
\begin{equation}
  g_i\ket{G}=\ket{G},
  \label{eq:GraphState}
\end{equation}
for all $i\in V$. The mutually commuting operators $g_i$ generate
the stabilizer group
\begin{equation}
  \cS=\Bigl\{g_1^{b_1}g_2^{b_2}\dots g_n^{b_n}\Bmid b_i=0,1\Bigr\}
  \label{eq:stabGroup}
\end{equation}
for $\ket{G}$.  Alternatively, an explicit expression of the graph state
$\ket{G}$ is
\begin{equation}
  \ket{G}=\prod_{\{i,j\}\in E}\CZ_{ij}\ket{+}^{\otimes n},
  \label{eq:GraphStateCZ}
\end{equation}
where the two-qubit control-$Z$ gates $\CZ_{ij}$, defined as
\begin{equation}
  \begin{aligned}
    \CZ_{ij}=&\ket{0}\bra{0}_i\otimes\I_j+\ket{1}\bra{1}_i\otimes Z_j\\
    =&\ket{0}\bra{0}_j\otimes\I_i+\ket{1}\bra{1}_j\otimes Z_i\\
    =&\I_i\otimes\I_j-2\ket{1}\bra{1}_i\otimes\ket{1}\bra{1}_j\\
  \end{aligned}
  \label{eq:CZ}
\end{equation}
are mutually commuting. Finally, it should be noted that Bell states
and GHZ states are, up to a local change of the basis, also graph states.

In Ref.~\cite{Pallister.etal2018}, the verification of graph states is
considered as a direct generalization for the verification of Bell states.
Let $\cS$ be the stabilizer group of $\ket{G}$, then one can construct
a verification strategy
\begin{equation}
  \Omega=\frac{1}{2^n-1}\sum_{g\in\cS,\,g\ne\I}P_{g}^+.
  \label{eq:stabStrategy}
\end{equation}
As in Eq.~\eqref{eq:GHZ}, $P_g^+=(\I^{\otimes n}+g)/2$ means that
the product of the measurement outcomes of the $n$ different parties
is one. The definition of the graph state implies the relation
\begin{equation}
  \ket{G}\bra{G}=\prod_{i=1}^n\frac{\I+g_i}{2}
  =\frac{1}{2^n}\sum_{g\in\cS}g,
\end{equation}
from which one obtains that
\begin{equation}
  \Omega=\frac{2^{n-1}}{2^n-1}\ket{G}\bra{G}
  +\frac{2^{n-1}-1}{2^n-1}\I.
\end{equation}
Thus, the spectral gap is given by
\begin{equation}
  \nu(\Omega)=\frac{2^{n-1}}{2^n-1}=\frac{1}{2}+\frac{1}{2^{n+1}-2}.
  \label{eq:stabSpec}
\end{equation}
Unlike the case of Bell states, this strategy is no longer optimal for the
general graph states. One example was already shown for the case of
GHZ states.  The spectral gap in Eq.~\eqref{eq:stabStrategy} is slightly
better than the verification strategy with two measurement settings in
Eq.~\eqref{eq:GHZ2Strategy}, but less efficient than the optimal strategy in
Eq.~\eqref{eq:GHZOptimalStrategy}.

In Ref.~\cite{Zhu.Hayashi2018}, Zhu and Hayashi put forward another efficient
method for verifying graph states. This method is closely related to the
coloring problem of the corresponding graphs, which also plays a central
role for efficient entanglement witnesses \cite{Toth.Guehne2005b}.
A (proper) coloring of a graph is a labeling of the graph’s vertices
with colors such that no two adjacent vertices have the same color.
A graph $G$ is called $m$-colorable if there exists a coloring with
$m$ colors and the smallest number of colors needed for coloring $G$
is called the chromatic number $\chi(G)$; some examples are shown in 
Fig.~\ref{fig:graph}.

Now, we can explain the coloring strategy for verifying graph states.
Let ${c:V\to \{1,2,\dots,m\}}$ be an $m$-coloring of a graph $G$.
Then, the following $m$ measurement settings are taken
\begin{equation}
  \bigotimes_{c(i)=\ell}X_i\otimes\bigotimes_{c(i)\ne \ell}Z_i
  \quad\text{for}~\ell=1,2,\dots,m.
  \label{eq:GraphColorMeasurement}
\end{equation}
For example, for the colored graph in Fig.~\ref{fig:graph}~(a), the measurement
settings read
\begin{equation}
  X\otimes Z\otimes X \otimes Z, \quad
  Z\otimes X\otimes Z \otimes X.
\end{equation}
The reason for choosing the measurement settings in
Eq.~\eqref{eq:GraphColorMeasurement} is that all the measurement outcomes of
the generators $g_i$ in Eq.~\eqref{eq:GraphStab}, or more precisely,
$\{P_{g_i}^+,\I-P_{g_i}^+\}$, can be inferred from the outcomes. Indeed, the
measurement outcome of $\{\Omega_\ell, \I-\Omega_\ell\}$ with
\begin{equation}
  \Omega_\ell=\prod_{c(i)=\ell}\frac{1+g_i}{2}
\end{equation}
can be inferred from the $\ell$-th measurement setting in
Eq.~\eqref{eq:GraphColorMeasurement}. Then, by choosing
\begin{equation}
  \Omega=\frac{1}{m}\sum_{\ell=1}^m\Omega_\ell,
  \label{eq:GraphColorStrategy}
\end{equation}
one can achieve the spectral gap
\begin{equation}
  \nu(\Omega)=\frac{1}{m}.
  \label{eq:Graph}
\end{equation}
With the coloring strategy, the largest spectral gap that can be achieved is
$1/\chi(G)$, where $\chi(G)$ is the chromatic number of the graph $G$.
But still, there exist some possible improvements and generalizations of the 
coloring strategy.

First, efficiency of the coloring strategy can be improved by the
LC operations \cite{VandenNest.etal2004}, which do not change the
entanglement properties of the graph state, but they change the graph
and hence the chromatic number.
In this way, the graph state can be transformed to a local unitary equivalent
graph state, which however may have a smaller chromatic number. For example,
the graph states in Fig.~\ref{fig:graph}~(b) and (d) are equivalent to those in
Fig.~\ref{fig:graph}~(a) and (c), respectively, but the chromatic numbers are
reduced from $3\to 2$ and $4\to 2$, respectively. This reduction can not only
simplify the measurement settings, but also improve the verification
efficiency.

Second, another way to improve the verification efficiency
is to take advantage of the so-called fractional coloring.
The corresponding strategy is called the fractional coloring
strategy (or cover strategy) \cite{Zhu.Hayashi2018}. In the
fractional coloring strategy, arbitrary independent sets instead
of disjoint independent sets are taken, and the largest achievable
spectral gap is $1/\chi_f(G)$, where $\chi_f(G)\le\chi(G)$ is called
the fractional chromatic number \cite{Scheinerman.Ullman1997}.

Third, as also shown in Ref.~\cite{Zhu.Hayashi2018}, both the coloring and
fractional coloring strategies can be directly generalized to the family of
so-called hypergraph states
\cite{Qu.etal2013,Rossi.etal2013,Guehne.etal2014b}.
Although the
stabilizer operators for hypergraph states are non-local as they contain
multi-qubit control-$Z$ operators, they can still be revealed by local
measurements.  This is because the measurement outcome of the $n$-qubit
control-$Z$ operator $C^{n-1}Z$ can be revealed by the local measurement
$Z^{\otimes n}$, mathematically speaking,
\begin{equation}
  \frac{\I^{\otimes n}-C^{n-1}Z}{2}
  =\bigotimes_{i=1}^n\frac{\I-Z_i}{2}.
\end{equation}
The coloring and fractional coloring of hypergraphs are also defined similarly
\cite{Scheinerman.Ullman1997}, from which the coloring and fractional coloring
strategies can be naturally generalized to the hypergraph states.

Recently, Dangniam \textit{et al.} put forward an algorithm for finding the
optimal verification strategies for graph states with Pauli measurements
\cite{Dangniam.etal2020}. The optimal verification strategies are based on the
so-called canonical test operators, and finding the optimal strategy can be
written as a linear program. Surprisingly, for all the graph states numerically
tested in Ref.~\cite{Dangniam.etal2020}, including all states with no more than
$7$ qubits, the maximal spectral gap is
\begin{equation}
  \nu(\Omega)=\frac{2}{3},
  \label{eq:GrapSpec}
\end{equation}
which is also optimal over all local or LOCC measurements. This motivates
the authors to conjecture that the upper bound $\nu(\Omega)=\frac{2}{3}$ can be
achieved for all graph states with Pauli measurements.

In addition, we would like to point out that there are also plenty of other
statistical methods for verifying graphs and hypergraph states
\cite{Flammia.Liu2011,Morimae.etal2017,Hayashi.Hajdusek2018,
Takeuchi.Morimae2018,Takeuchi.etal2019}; see Ref.~\cite{Zhu.Hayashi2018} for
a detailed comparison of these methods.

\subsubsection{General pure two-qubit states}%
All the above strategies are related to the stabilizer formalism. For
general states, it is however difficult to construct an optimal verification
strategy.  The only known optimal result is for two-qubit pure states with
local projective measurements. Without loss of generality, we can assume that
the general (not separable or maximally entangled) pure quantum state is of the
form
\begin{equation}
  \ket{\psi}=\cos\theta\ket{00}+\sin\theta\ket{11},
  \label{eq:2qubit}
\end{equation}
where $0<\theta<\pi/4$.

In Ref.~\cite{Pallister.etal2018}, Pallister \textit{et al.} showed that the
optimal strategy for verifying the state in Eq.~\eqref{eq:2qubit} is
\begin{equation}
  \Omega=\alpha(\theta)P_{ZZ}^++\frac{1-\alpha(\theta)}{3}\sum_{\ell=1}^3
  (\I-\ket{u_\ell}\bra{u_\ell}\otimes\ket{v_\ell}\bra{v_\ell}),
  \label{eq:2qubitStrategy}
\end{equation}
where $\alpha(\theta)=\frac{2-\sin(2\theta)}{4+\sin(2\theta)}$ and
$\ket{u_\ell}\ket{v_\ell}$ are some states such that
$\bra{u_\ell}\braket{v_\ell}{\psi}=0$. The optimal spectral gap is given by
\begin{equation}
  \nu(\Omega)=\frac{1}{2+\sin\theta\cos\theta}.
  \label{eq:2qubitSpec}
\end{equation}

\begin{figure}[t]
  \centering
  \includegraphics[width=.45\textwidth]{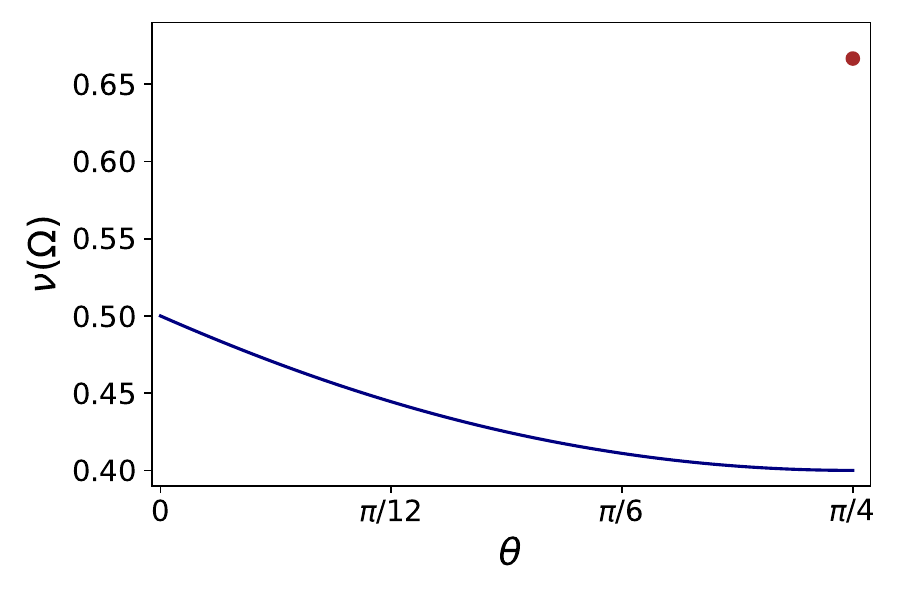}
  \caption{%
    Optimal verification of $\ket{\psi}=\cos\theta\ket{00}+\sin\theta\ket{11}$
    with local projective measurements. When $0<\theta<\pi/4$, the spectral gap
    is shown in Eq.~\eqref{eq:2qubitSpec}, which satisfies that
    $\lim_{\theta\to\pi/4}\nu(\Omega)=2/5$. However, when $\theta=\pi/4$, the
    optimal gap is $2/3$, as shown in Eq.~\eqref{eq:BellSpec}.
  }
  \label{fig:pcl}
\end{figure}

As shown in Fig.~\ref{fig:pcl}, the optimal spectral gap is not continuous at
$\theta=\pi/4$, which suggests that there may exist methods to improve
the verification efficiency. Indeed, unlike in the case of Bell states, the
strategy in Eq.~\eqref{eq:2qubitStrategy} is no longer optimal if local POVMs
are considered, at least for some $0<\theta<\pi/4$.
An easy way to see this is to perform a local filtering to make the state
maximally entangled and then perform the optimal verification for the resulted
state. In this way, the verification efficiency will be continuous at
$\theta=\pi/4$. However, finding the optimal strategy with local POVMs is still
an open problem, even for two-qubit states. Another method for improving the
verification efficiency is to take advantage of LOCC measurements, which is the
main topic of the next subsection.

The first experimental implementation of QSV on two-qubit entangled 
photonic states with local measurements was reported in Ref.~\cite{Zhang.etal2020}.
Considering the robustness to practical imperfections, the actual realization 
was relying on the Chernoff-Hoeffding bound in Eq.~\eqref{eq:CHbound}.
The inverse proportionality between the estimated infidelity and the number 
of samples was clearly demonstrated by all the tested states. For the estimated 
infidelity of, say $\epsilon=0.01$, the confidence level rapidly approaches 
near-unity within $1000$ number of samples. Moreover, to show the scalability 
of the QSV methodology, a four-qubit GHZ state was also verified.

\subsection{QSV with LOCC measurements}

\begin{figure}[t]
  \centering
  \includegraphics[width=.45\textwidth]{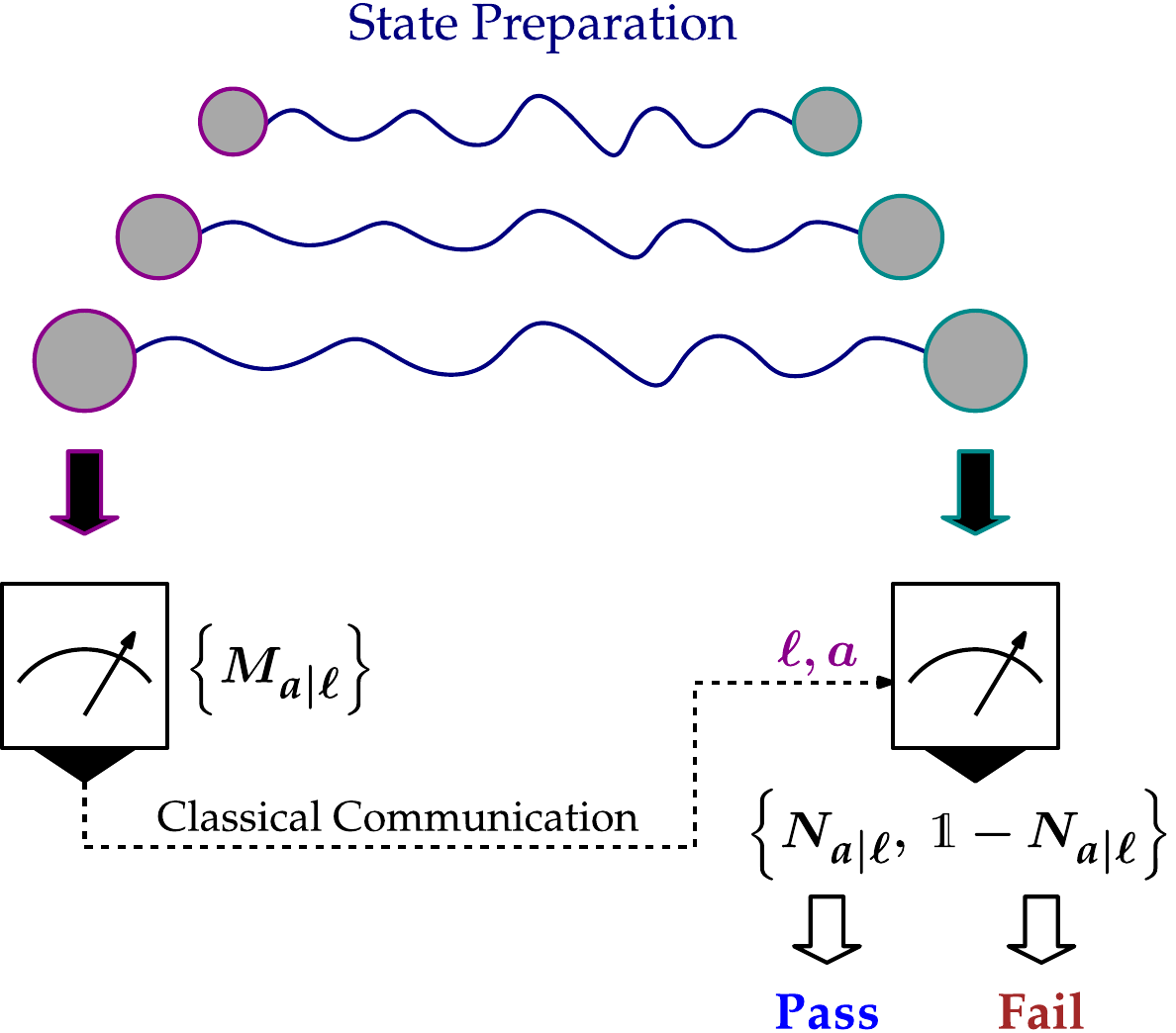}
  \caption{%
    QSV with one-way LOCC measurements. In the $k$-th run, Alice randomly
    chooses a measurement $\cM_\ell$ with probability $p_\ell$. After
    performing the measurement $\cM_\ell$, Alice tells Bob via classical
    communication the measurement setting $\ell$ and the outcome $a$, based on
    which Bob performs the pass or fail test $\{N_{a|\ell},\I-N_{a|\ell}\}$.
  }
  \label{fig:locc}
\end{figure}

QSV with LOCC measurements was proposed independently in
Refs.~\cite{Yu.etal2019,Wang.Hayashi2019,Li.etal2019}. LOCC is a method in
quantum information theory where the local operations are assisted by
classical communication between the parties \cite{Horodecki.etal2009}. The set
of general LOCC measurements is complicated due to the unbounded number of
communication rounds \cite{Kleinmann.etal2011,Chitambar2011}. As a result, most
of the works on QSV with LOCC measurements focus on the simplified scenario of
one-round communication.

We start with the analysis of the one-way LOCC strategy for two parties, Alice
and Bob, as illustrated in Fig.~\ref{fig:locc}. In this case, Alice first
performs a random measurement $\cM_\ell=\{M_{a|\ell}\}_a$ with probability
$p_\ell$ on her subsystem, and sends to Bob the measurement setting $\ell$ and
the outcome $a$, based on which Bob performs a measurement
$\{N_{a|\ell},\I-N_{a|\ell}\}$ on his subsystem.  The state $\sigma_k$ passes
the test if and only if Bob obtains the outcome corresponding to $N_{a|\ell}$.
Thus, the one-way LOCC strategy $\Omega^\rightarrow$ is of the form
\begin{equation}
  \Omega^\rightarrow=\sum_{\ell=1}^np_\ell\Omega_\ell^\rightarrow,\quad
  \Omega_\ell^\rightarrow=\sum_aM_{a|\ell}\otimes N_{a|\ell}.
  \label{eq:oneWayStrategy}
\end{equation}

Without loss of generality, one can assume that the measurement operators
$M_{a|\ell}$ are rank-one. If this is not the case, one can always perform
a finer measurement by measuring the spectral decomposition. Then, Alice's
measurement $\cM_\ell$ would cause Bob's subsystem to collapse to the pure
state
\begin{equation}
  {P_{a|\ell}=\frac{\Tr_A\big[(M_{a|\ell}\otimes\I)\ket{\psi}\bra{\psi}\big]}
  {\Tr\big[(M_{a|\ell}\otimes\I)\ket{\psi}\bra{\psi}\big]}},
  \label{eq:postmeasurement}
\end{equation}
where $a$ is the corresponding measurement outcome. For a fixed $\cM_\ell$, the
optimal strategy for Bob is to take
\begin{equation}
  N_{a|\ell}=P_{a|\ell}.
  \label{eq:semioptimal}
\end{equation}
Thus, a strategy of this form is called semi-optimal, as it is optimal
concerning Bob's side. From the
definition, one can easily prove the following necessary conditions for
$\Omega^\rightarrow$ being semi-optimal
\begin{equation}
  \Omega^\rightarrow\in\Sep,\quad
  \Tr_B(\Omega^\rightarrow)=\I,\quad
  \bra{\psi}\Omega^\rightarrow\ket{\psi}=1,
  \label{eq:oneWayConstraints}
\end{equation}
where $\Sep$ is the set of unnormalized separable states, i.e.,
\begin{equation}
  \Sep:=\biggl\{\sum_{i}X_i^A\otimes Y_i^B\Bmid
  ~X_i^A\ge 0,~Y_i^B\ge 0\biggr\}.
  \label{eq:Sep}
\end{equation}
On the other hand, if $\Omega^\rightarrow$ is of the form in
Eq.~\eqref{eq:oneWayConstraints}, one can also verify that it has
a decomposition of the form in Eq.~\eqref{eq:oneWayStrategy}
\cite{Yu.etal2019}. Thus, constructing the optimal one-way LOCC strategy can be
written as the following convex optimization problem
\begin{equation}
  \begin{aligned}
    &\maxover[\Omega^\rightarrow] \quad &&\nu(\Omega^\rightarrow)\\
    &\subto                       &&\Omega^\rightarrow\in\Sep,\\
    &                             &&\Tr_B(\Omega^\rightarrow)=\I,\\
    &                             &&\bra{\psi}\Omega^\rightarrow\ket{\psi}=1.
  \end{aligned}
  \label{eq:oneWayCvx}
\end{equation}

Let us continue to discuss the case of one-round two-way LOCC strategies. In
this case, Alice and Bob use shared randomness to decide who first performs the
measurement.  After the measurement, the measurement outcome is sent to the
other party.  The receiver then performs the pass or fail test
according to the received measurement outcome. Further, up to
some local unitary operation, a general bipartite pure state can be written
in the Schmidt decomposition
\begin{equation}
  \ket{\psi}=\sum_{\alpha=1}^{d}\lambda_\alpha\ket{\alpha\alpha},
  \label{eq:generalBipartite}
\end{equation}
where $\lambda_1\ge\lambda_2\ge\dots\ge\lambda_d\ge0$.

Thanks to the permutation
symmetry of $\ket{\psi}$ in Eq.~\eqref{eq:generalBipartite}, the optimization
in this setting can be easily simplified.  Let $V$ be the SWAP operator, i.e.,
\begin{equation}
  V\ket{\alpha}\ket{\beta}=\ket{\beta}\ket{\alpha}
  ~~~\text{for all}~\alpha,\beta=1,2,\dots,d,
  \label{eq:swap}
\end{equation}
then we have $V\ket{\psi}=\ket{\psi}$. This indicates that, if $\Omega$ is
a two-way LOCC strategy, so is $(\Omega+V\Omega V^\dagger)/2$.  Furthermore,
one can easily show that
\begin{equation}
  \nu\Bigl[\tfrac{1}{2}(\Omega+V\Omega V^\dagger)\Bigr]\ge
  \tfrac{1}{2}\Bigl[\nu(\Omega)+\nu(V\Omega V^\dagger)\Bigr]=\nu(\Omega).
  \label{eq:swapSymmetry}
\end{equation}
Hence, one can focus on the two-way LOCC strategies that are invariant under
the SWAP operation. When restricted to the one-round case, constructing the
optimal strategy $\Omega^\leftrightarrow$ can be written as
\begin{equation}
  \begin{aligned}
    &\maxover[\Omega^\rightarrow] \quad &&\nu\Bigl[\tfrac{1}{2}
			(\Omega^\rightarrow+\Omega^\leftarrow)\Bigr]\\
    &\subto                       &&\Omega^\rightarrow\in\Sep,\\
    &                             &&\Tr_B(\Omega^\rightarrow)=\I,\\
    &                             &&\bra{\psi}\Omega^\rightarrow\ket{\psi}=1,
  \end{aligned}
  \label{eq:twoWayCvx}
\end{equation}
where $\Omega^\rightarrow$ is a one-way LOCC strategy,
$\Omega^\leftarrow=V\Omega^\rightarrow V^\dagger$, and the constraints are from
Eq.~\eqref{eq:oneWayCvx}.

\begin{figure}[t]
  \centering
  \includegraphics[width=.45\textwidth]{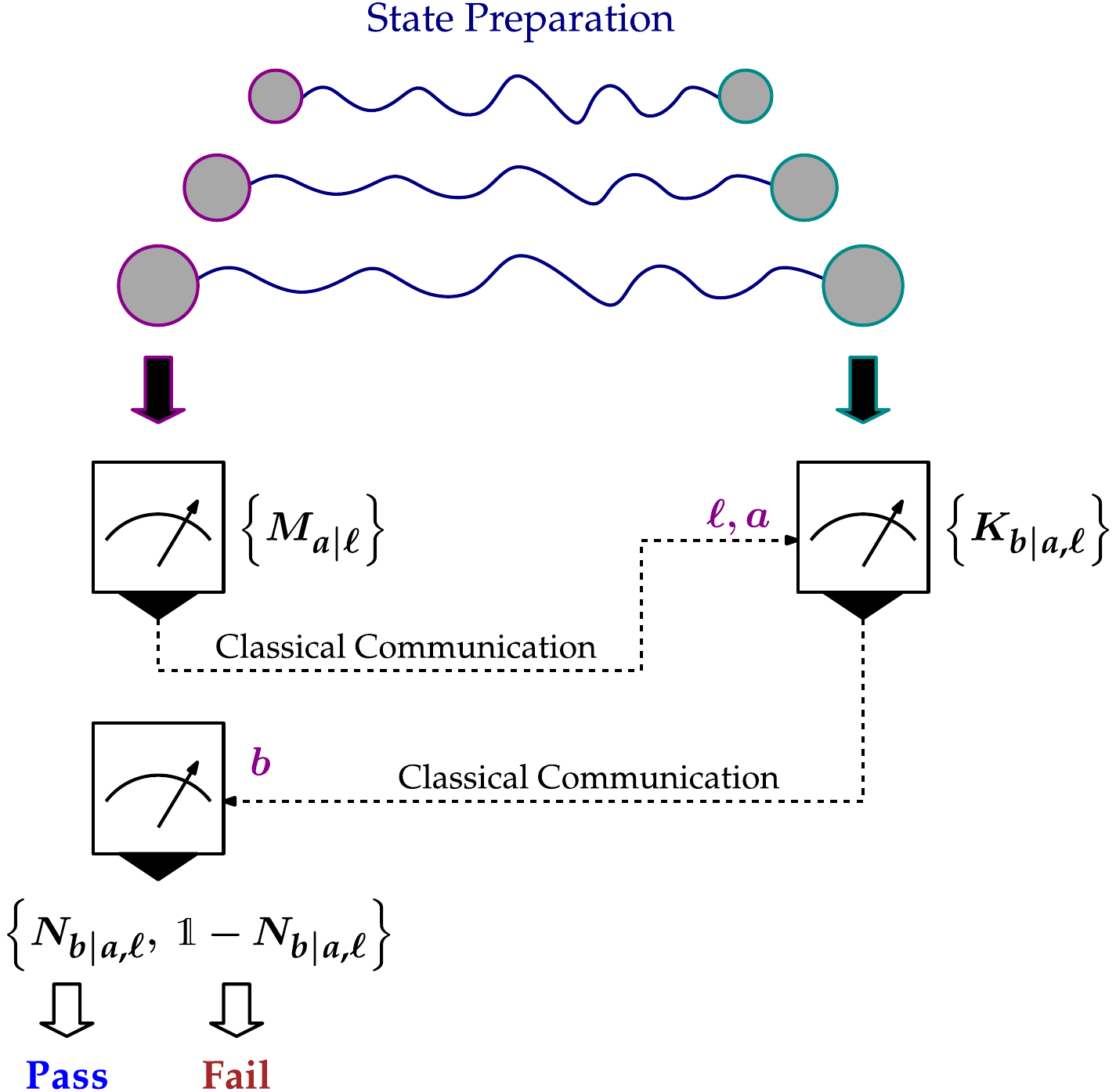}
  \caption{%
    QSV with many-round LOCC measurements. In the protocol, nondestructive
    measurements are performed and many-round communication is involved between
    Alice and Bob.
  }
  \label{fig:locc2}
\end{figure}

At last, we note that it is possible to further improve the verification
efficiency by considering many-round communication; see Fig.~\ref{fig:locc2}.
A concrete example will be shown for two-qubit pure states below.

\subsubsection{General two-qubit pure states}%

As in Eq.~\eqref{eq:2qubit}, we write the general two-qubit
entangled pure state as $\ket{\psi}=\cos\theta\ket{00}+\sin\theta\ket{11}$
with $0<\theta\le\pi/4$.  Then the optimization problems in
Eqs.~(\ref{eq:oneWayCvx},\,\ref{eq:twoWayCvx}) are directly solvable
because the positive partial transpose (PPT) criterion provides a
necessary and sufficient condition for the separability problem
\cite{Peres1996,Horodecki.etal1996}. Indeed, the optimization
problems can be solved analytically by taking advantage of the
symmetry.

Contrary to the maximally entangled state, a general bipartite pure
state has no longer the $U\otimes U^*$ symmetry as in Eq.~\eqref{eq:MESsymmetry}.
However, a restricted symmetry still holds where the $U$ are constrained
to be the diagonal unitary matrices \cite{Pallister.etal2018}. This is
also equivalent to the discrete group $\cG$ generated by the following
phase gate
\begin{equation}
  g_0=
  \begin{bmatrix}
    1 & 0\\
    0 & \mi
  \end{bmatrix}
  \otimes
  \begin{bmatrix}
    1 & 0\\
    0 & -\mi
  \end{bmatrix},
  \label{eq:phaseGate}
\end{equation}
which also plays an important role in constructing the optimal
verification strategy \cite{Yu.etal2019}. One can easily verify
that $\cG=\{\I,g_0,g_0^2,g_0^3\}$ and
\begin{equation}
  \nu\biggl(\sum_{g\in\cG}g\Omega g^{\dagger}\biggr)\ge\frac{1}{\abs{\cG}}\sum_{g\in\cG}
  \nu\Bigl(g\Omega g^\dagger\Bigr)=\nu(\Omega).
  \label{eq:loccSymmetrize}
\end{equation}
Thus, without loss of generality, one can assume that the optimal $\Omega$ is
invariant under $\cG$. Then, by taking advantage of the PPT criterion,
one obtains the optimal verification strategy by solving the optimization in
Eq.~\eqref{eq:oneWayCvx}, and the spectral gap is given by
\begin{equation}
  \max_{\Omega^\rightarrow}\nu(\Omega^\rightarrow)
  =\frac{1}{1+\cos^2\theta}.
  \label{eq:oneWayQubitSpec}
\end{equation}

As shown in Refs.~\cite{Yu.etal2019,Wang.Hayashi2019,Li.etal2019}, this optimal
spectral gap can be achieved already with the following projective
measurements
\begin{equation}
\Omega^{\rightarrow}=
\frac{1}{(1+\cos^2\theta)}
\big[(\cos^2\theta)P_{ZZ}^+
  +\frac{1}{2}X_\psi^{\rightarrow}
  +\frac{1}{2}Y_\psi^{\to}
\big],
\label{eq:oneWayQubitPVM}
\end{equation}
where
\begin{equation}
  \begin{aligned}
    P_{ZZ}^+&=\ket{0}\bra{0}\otimes\ket{0}\bra{0}
    +\ket{1}\bra{1}\otimes\ket{1}\bra{1},\\
    X_\psi^{\rightarrow}&=\ket{\varphi_0}\bra{\varphi_0}
    +\ket{\varphi_2}\bra{\varphi_2},\\
    Y_\psi^{\rightarrow}&=\ket{\varphi_1}\bra{\varphi_1}
    +\ket{\varphi_3}\bra{\varphi_3},
  \end{aligned}
  \label{eq:defXYZ}
\end{equation}
with
$\ket{\varphi_0}=\frac{1}{\sqrt{2}}(\ket{0}+\ket{1})
\otimes(\cos\theta\ket{0}+\sin\theta\ket{1})$ and
$\ket{\varphi_k}=g_0^k\ket{\varphi_0}$ for $k=1,2,3$.

Similarly, for the case of one-round two-way LOCC measurements, the optimal
spectral gap, i.e., the solution of the optimization in
Eq.~\eqref{eq:twoWayCvx}, is given by
\begin{equation}
  \max_{\Omega^\leftrightarrow}\nu(\Omega^\leftrightarrow)
  =\frac{2}{3}.
  \label{eq:twoWayQubitSpec}
\end{equation}
which can also be achieved by projective measurements
\cite{Yu.etal2019,Wang.Hayashi2019,Li.etal2019}
\begin{equation}
    \Omega^{\leftrightarrow}=\frac{1}{3}P_{ZZ}^++\frac{1}{6}X_\psi^{\rightarrow}
    +\frac{1}{6}X_\psi^{\leftarrow}+\frac{1}{6}Y_\psi^{\rightarrow}
    +\frac{1}{6}Y_\psi^{\leftarrow},
  \label{eq:twoWayQubitPVM}
\end{equation}
where $P_{ZZ}^+$, $X_\psi^{\rightarrow}$, and $Y_\psi^{\rightarrow}$ are
defined as in Eq.~\eqref{eq:defXYZ}, and
$X_\psi^{\leftarrow}=VX_\psi^{\rightarrow}V^{\dagger}$ and
$Y_\psi^{\leftarrow}=VY_\psi^{\rightarrow}V^{\dagger}$, with
$V$ being the SWAP operator.

In addition, it is also possible to improve the verification efficiency further
by taking advantage of many-round communication \cite{Wang.Hayashi2019}.
The building block $\{X_{\psi,\eta}^{A\Leftrightarrow B},
\I-X_{\psi,\eta}^{A\Leftrightarrow B}\}$ is the following measurement
procedure:
\begin{enumerate}[left=0pt]
  \item
    Alice first performs a nondestructive POVM $\{M_0:=\eta\ket{0}\bra{0},
    M_1:=\I-\eta\ket{0}\bra{0}\}$ and sends the measurement outcome $0$ or $1$
    to Bob.
  \item
    If the measurement outcome of Alice is zero, Bob performs the measurement
    $Z$ on his subsystem and accepts (rejects) the test if he obtains the
    outcome $+1$~($-1$). If the measurement outcome of Alice is one, Bob
    performs the measurement $X$ on his subsystem and sends the measurement
    outcome $+1$ or $-1$ back to Alice.
  \item
    Based on the measurement outcome of Bob, Alice performs the corresponding
    test to check whether the subsystem is in the post-measurement state
    $\ket{v_+}\bra{v_+}$ or $\ket{v_-}\bra{v_-}$, which is defined similarly
    as the in semi-optimal strategy in 
    Eqs.~(\ref{eq:postmeasurement},\,\ref{eq:semioptimal}).
\end{enumerate}
Mathematically, this procedure (for the pass instances) can be written as
\begin{widetext}
\begin{equation}
  \begin{aligned}
    \rho\rightarrow&
    \Bigl[\sqrt{M_0}\otimes\ket{0}\bra{0}\Bigr]\rho
    \Bigl[\sqrt{M_0}\otimes\ket{0}\bra{0}\Bigr]\\
    &+\Bigl[\Bigl(\ket{v_+}\bra{v_+}\sqrt{M_1}\Bigr)\otimes
    \ket{+}\bra{+}\Bigr]\rho
    \Bigl[\Bigl(\sqrt{M_1}\ket{v_+}\bra{v_+}\Bigr)
    \otimes\ket{+}\bra{+}\Bigr]\\
    &+\Bigl[\Bigl(\ket{v_-}\bra{v_-}\sqrt{M_1}\Bigr)
    \otimes\ket{-}\bra{-}\Bigr]\rho
    \Bigl[\Bigl(\sqrt{M_1}\ket{v_-}\bra{v_-}\Bigr)
    \otimes\ket{-}\bra{-}\Bigr],
  \end{aligned}
  \label{eq:gg}
\end{equation}
where
\begin{equation}
  \ket{v_\pm}\bra{v_\pm}
  =\frac{\Tr_B\bigl[\bigl(\sqrt{M_1}\otimes
      \ket{\pm}\bra{\pm}\bigr)\ket{\psi}\bra{\psi}
  \bigl(\sqrt{M_1}\otimes\ket{\pm}\bra{\pm}\bigr)\bigr]}
  {\Tr\bigl[\bigl(\sqrt{M_1}\otimes\ket{\pm}\bra{\pm}\bigr)
      \ket{\psi}\bra{\psi}
  \bigl(\sqrt{M_1}\otimes\ket{\pm}\bra{\pm}\bigr)\bigr]}.
  \label{eq:vpm}
\end{equation}
\end{widetext}
Accordingly, $X_{\psi,\eta}^{A\Leftrightarrow B}$ reads
\begin{align}
    X_{\psi,\eta}^{A\Leftrightarrow B}
    =&M_0\otimes\ket{0}\bra{0}
    +\Bigl(\sqrt{M_1}\ket{v_+}\bra{v_+}\sqrt{M_1}\Bigr)
    \otimes\ket{+}\bra{+}\nonumber\\
    &+\Bigl(\sqrt{M_1}\ket{v_-}\bra{v_-}\sqrt{M_1}\Bigr)
    \otimes\ket{-}\bra{-}.
    \label{eq:XpsiABA}
\end{align}

\begin{figure}[t!!]
  \includegraphics[width=.45\textwidth]{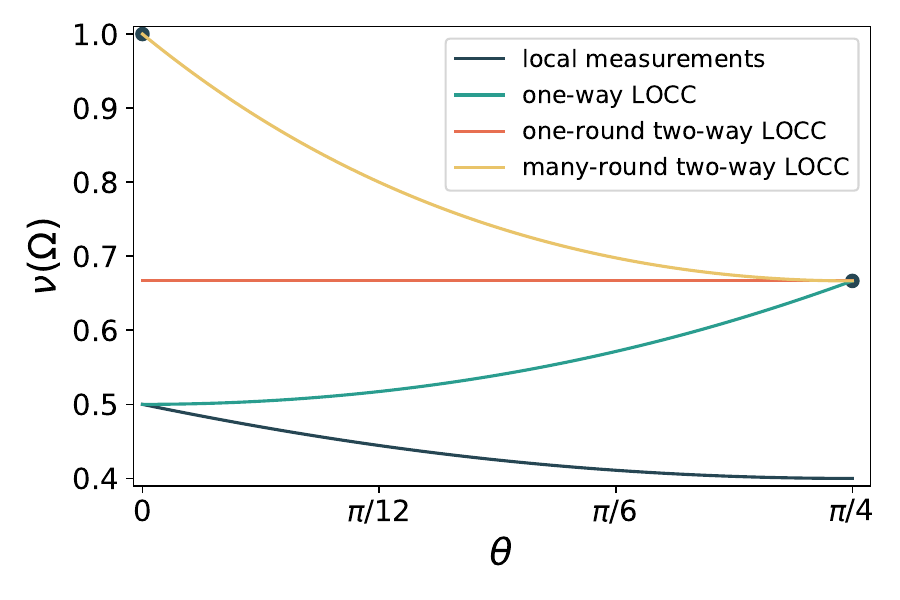}
  \caption{%
    Optimal values of $\nu(\Omega)$ with different verification strategies for
    the two-qubit entangled pure state
    $\ket{\psi}=\cos{\theta}\ket{00}+\sin{\theta}\ket{11}$ with
    $0\le\theta\le{\pi}/{4}$. Note that when $\theta=0$ or $\theta={\pi}/{4}$,
    all strategies give the same optimal spectral gap $\nu(\Omega)=1$ or
    $\nu(\Omega)=2/3$.
  }
  \label{fig:comparison}
\end{figure}

By taking advantage of the symmetry, one can construct similar procedures
$Y_{\psi,\eta}^{A\Leftrightarrow B}$, in which the measurement $X$ is replaced
by $Y$, as well as $X_{\psi,\eta}^{B\Leftrightarrow A}$ and
$Y_{\psi,\eta}^{B\Leftrightarrow A}$, in which Bob, instead of Alice, starts
the measurement. In Ref.~\cite{Wang.Hayashi2019}, Wang and Hayashi proved that
the optimal verification strategy with many-round communication can be achieved
by
\begin{equation}
    \Omega^{\Leftrightarrow}=pP_{ZZ}^++\frac{1-p}{4}
    \Big(X_{\psi,\eta}^{A\Leftrightarrow B}
      +X_{\psi,\eta}^{B\Leftrightarrow A}
      +Y_{\psi,\eta}^{A\Leftrightarrow B}
      +Y_{\psi,\eta}^{B\Leftrightarrow A}
    \Big),
  \label{eq:multiRoundQubitPVM}
\end{equation}
where $\eta=1-{\tan\theta}$ and
$p=(\sin^2\theta)/(1+\sin\theta\cos\theta)$.
The optimal spectral gap is
\begin{equation}
  \nu(\Omega^\Leftrightarrow)=\frac{1}{1+\sin\theta\cos\theta}.
  \label{eq:manyRoundQubitSpec}
\end{equation}
Remarkably, this strategy is also optimal for the case that infinite rounds
of classical communication are allowed. The comparison of the verification 
efficiency (spectral gap) of different local and LOCC strategies is illustrated 
in Fig.~\ref{fig:comparison}.

Using photonic systems, both the experiments reported in 
Refs.~\cite{Jiang.etal2020, Zhang.etal2020b} have successfully 
demonstrated the enhanced QSV strategies using classical communication. 
For instance, it was shown in Ref.~\cite{Zhang.etal2020b} that only $60\%$ 
of the measurements are required to achieve a certain value of precision 
as compared to the optimal strategy with local projective measurements.
Hence, the experimental 
results clearly indicate that classical communication can significantly 
enhance the performance of QSV, and lead to an 
efficiency that further approaches the globally optimal bound.

\subsubsection{General bipartite pure states}%
We move on to discuss the verification of general bipartite pure states with
LOCC measurements. As in Eq.~\eqref{eq:generalBipartite}, we write the
bipartite pure state as
$\ket{\psi}=\sum_{\alpha=1}^{d}\lambda_\alpha\ket{\alpha\alpha}$, where
$\lambda_1\ge\lambda_2\ge\dots\ge\lambda_d$.

To construct the optimal LOCC strategy, a crucial ingredient is the group $\cG$
generated by
\begin{align}
  \label{eq:phaseGeneral}
  &g_\alpha=\Phi_\alpha\otimes\Phi_\alpha^\dagger,\quad \alpha=1,2,\dots,d,\\
  &\Phi_\alpha\ket{\beta}=
  \begin{cases}
    \mi\ket{\beta} \quad &\text{when}~\alpha=\beta,\\
    \ket{\beta}  \quad &\text{when}~\alpha\ne\beta,
  \end{cases}
\end{align}
which generalizes Eq.~\eqref{eq:phaseGate} to the two-qudit case. Similar to
Eq.~\eqref{eq:loccSymmetrize}, we can also assume the optimal $\Omega$ is
invariant under $\cG$.

The main difference between two-qudit and two-qubit states is that the PPT
criterion is no longer sufficient for characterizing the separability when $d\ge
3$ \cite{Horodecki.etal1996}. Hence, by replacing $\Omega^\rightarrow\in\Sep$
with $\Omega^\rightarrow\ge 0$ and $(\Omega^\rightarrow)^{T_B}\ge 0$,
Eqs.~(\ref{eq:oneWayCvx},\,\ref{eq:twoWayCvx}) only give us relaxations of the
original optimization problems, and the solutions of the relaxed problems only
result in upper bounds of the optimal $\nu(\Omega^\rightarrow)$ and
$\nu(\Omega^\leftrightarrow)$.

For the one-way LOCC strategy, the relaxed optimization problem can be solved
analytically \cite{Yu.etal2019}, which gives the upper bound
\begin{equation}
  \max_{\Omega^\rightarrow}\nu(\Omega^\rightarrow)\le\frac{1}{1+\lambda_1^2},
  \label{eq:oneWayQuditSpec}
\end{equation}
for all $d\ge 2$. Fortunately, this upper bound can be achieved with one-way
LOCC measurements, or even with projective ones. The corresponding verification
strategy reads
\begin{equation}
  \Omega^\rightarrow=\omega P_{ZZ}+\frac{1-\omega}{\abs{\cG}}
  \sum_{g\in\cG}gX_\psi^\rightarrow g^\dagger,
  \label{eq:oneWayQuditStrategy}
\end{equation}
where
\begin{equation}
  \begin{aligned}
    &P_{ZZ}=\sum_{\alpha=1}^d\ket{\alpha}\bra{\alpha}\otimes
    \ket{\alpha}\bra{\alpha},
    ~X_\psi^\rightarrow=\sum_{\alpha=1}^d
    \ket{f_\alpha}\bra{f_\alpha}
    \otimes\ket{\phi_\alpha}\bra{\phi_\alpha},\\
    &\ket{f_\alpha}=\frac{1}{\sqrt{d}}\sum_{\beta=1}^d
    \zeta_d^{\alpha\beta}\ket{\beta},
    \quad\ket{\phi_\alpha}=\sum_{\beta=1}^d\zeta_d^{-\alpha\beta}\lambda_\beta\ket{\beta},
  \end{aligned}
  \label{eq:fourierBasis}
\end{equation}
with $\zeta_d=\Exp{\frac{2\pi\mi}{d}}$ and
$\omega=\lambda_1^2/(1+\lambda_1^2)$. In experiments, the above strategy can be
easily implemented with the random measurement in the computational basis
$\{\ket{\alpha}\}_{\alpha=1}^d$ or in the Fourier basis
$\{\ket{f_\alpha}\}_{\alpha=1}^d$ accompanied by random phase shifts from
$\cG$. Especially, when $\ket{\psi}$ is maximally entangled,
$\{\ket{\phi_\alpha}\}_{\alpha=1}^d$ forms an orthogonal basis. Hence,
Eq.~\eqref{eq:fourierBasis} gives an alternative optimal strategy for verifying
maximally entangled states with local projective measurements.

For one-round two-way LOCC strategies, the verification efficiency can also be
improved by averaging $\Omega^\rightarrow$ and its swap $\Omega^\leftarrow$.
Specifically, one can achieve the spectral gap
\begin{equation}
  \nu(\Omega^\leftrightarrow)=\nu\Bigl[\tfrac{1}{2}(\Omega^\rightarrow
  +\Omega^\leftarrow)\Bigr]=\frac{1}{1+\lambda^2},
  \label{eq:twoWayQuditSpec}
\end{equation}
when $\Omega^\rightarrow$ is of the form in Eq.~\eqref{eq:oneWayQuditStrategy}
with
\begin{equation}
  \omega=\frac{\lambda^2}{1+\lambda^2},\quad
  \lambda^2=\frac1{2}\bigl(\lambda_1^2+\lambda_2^2\bigr).
\end{equation}
Unlike the two-qubit case, this one-round two-way LOCC strategy is only
nearly optimal for general bipartite states \cite{Yu.etal2019}.

Remarkably, as shown in Ref.~\cite{Li.etal2019}, the optimal and near-optimal
spectral gap in Eqs.~(\ref{eq:oneWayQuditSpec},\,\ref{eq:twoWayQuditSpec})
can also be achieved by taking advantage of the complete set of MUBs or with
$2$-designs, which are essentially different implementations (decompositions)
of the same verification operators $\Omega^\rightarrow$ or
$\Omega^\leftrightarrow$ in
Eqs.~(\ref{eq:oneWayQuditStrategy},\,\ref{eq:twoWayQuditSpec}).

\begin{figure}
  \includegraphics[width=.48\textwidth]{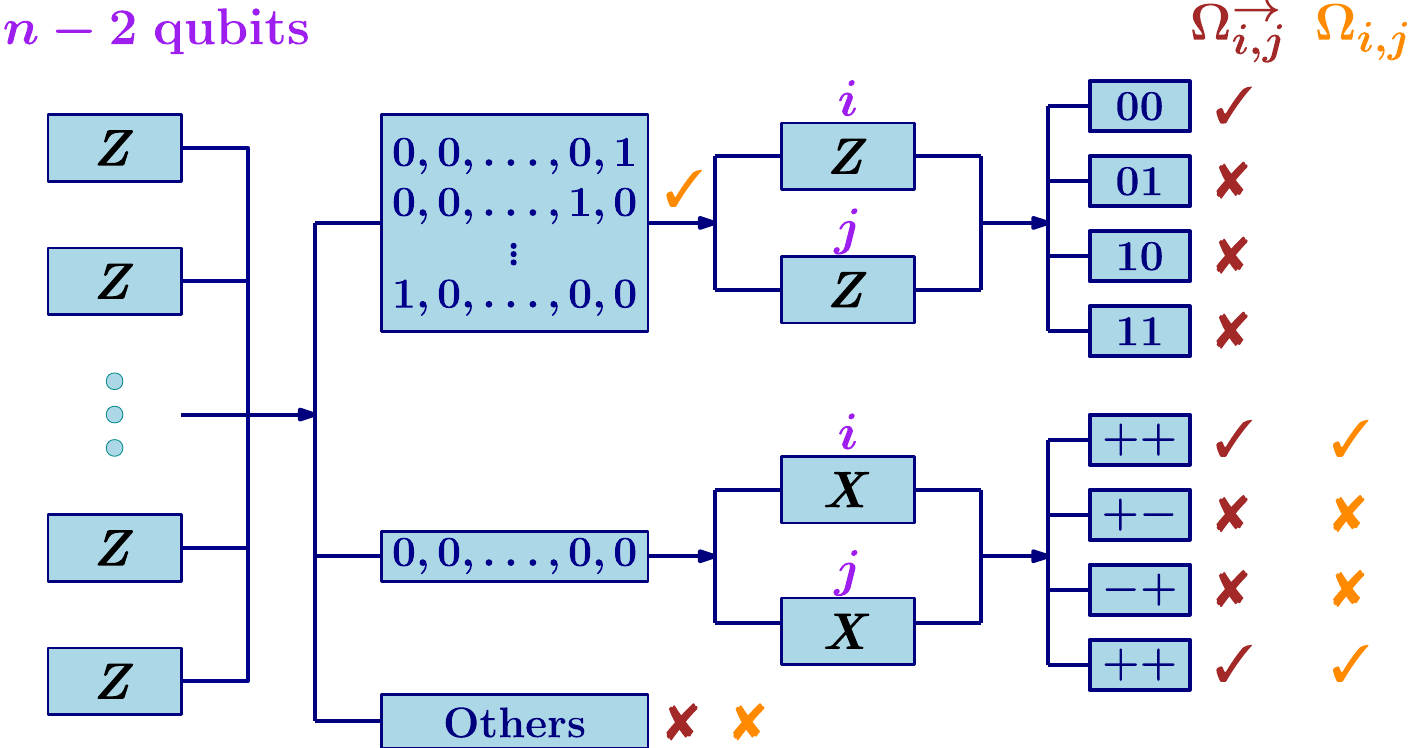}
  \caption{%
    $\Omega_{i,j}^\rightarrow$ and $\Omega_{i,j}$ for the verification of
    $\ket{W_n}$. For both $\Omega_{i,j}^\rightarrow$ and $\Omega_{i,j}$,
    the measurement $Z^{\otimes(n-2)}$ is performed on all but the $i$-th and
    $j$-th qubits. The difference is that for $\Omega_{i,j}$ the measurement on
    the $i$-th and $j$-th qubits is always $X\otimes X$, but for
    $\Omega_{i,j}^\rightarrow$ the measurement on the $i$-th and $j$-th qubits
    depends on the measurement outcome on the previous $n-2$ qubits.
  }
  \label{fig:vw}
\end{figure}
%

\subsubsection{$W$ states and Dicke states}%
The LOCC strategies can be easily generalized for verifying multi-party
states. Especially, the first verification strategy for non-stabilizer states
was constructed by Liu \textit{et al.} in Ref.~\cite{Liu.etal2019} for $W$
states and Dicke states.

We take $W$ states to illustrate the idea for verifying multi-party states
with LOCC measurements. The $n$-qubit $W$ state
\cite{Duer.etal2000,Haeffner.etal2005} is defined as
\begin{equation}
  \ket{W_n}=\frac1{\sqrt{n}}(\ket{10\dots0}
  +\ket{01\dots0}+\cdots+\ket{00\dots1}).
  \label{eq:Wstate}
\end{equation}
The verification strategy for $\ket{W_n}$ is based on the following two
observations (which have also been used to derive non-locality arguments
\cite{Heaney.etal2011}): First, the state $\ket{W_n}$ is symmetric under
permutations. Second, if the verifier performs the measurement
$Z^{\otimes(n-2)}$ on the first $n-2$ qubits, then outcome $1$
can appear at most once; otherwise, the original state cannot be $\ket{W_n}$.
If outcome $1$ appears, then the post-measurement state of parties $n-1$ and
$n$ will be $\ket{00}$, which can be verified easily by $Z\otimes Z$
measurement on these two qubits; if outcome $1$ does not appear, then the
post-measurement state will be the Bell state $(\ket{01}+\ket{10})/\sqrt{2}$,
which can be verified with local measurements as shown in
Eq.~\eqref{eq:BellStrategy}. By taking advantage of the permutation symmetry,
it can be shown that the tests based on $Y\otimes Y$ and $Z\otimes Z$
measurements in Eq.~\eqref{eq:BellStrategy} can be dropped.  More precisely,
$\ket{W_n}$ can be verified efficiently using the strategy
\begin{equation}
  \Omega^\leftrightarrow=\frac{2}{n(n-1)}\sum_{i<j}\Omega_{i,j}^\rightarrow
  \label{eq:WStrategy}
\end{equation}
with
\begin{equation}
  \Omega_{i,j}^{\rightarrow}
  =\bar{\cZ}_{i,j}^1(Z_i^{+}Z_j^{+})
  +\bar{\cZ}_{i,j}^0(XX)_{i,j}^{+},
  \label{eq:WStrategyComp}
\end{equation}
where $\bar{\cZ}_{i,j}^k$ denotes that exactly $k$ excitations are detected
when the measurement $Z^{\otimes(n-2)}$ is performed on all but the $i$-th and
$j$-th qubits; see Fig.~\ref{fig:vw}. The corresponding spectral gap reads
\begin{equation}
  \nu(\Omega^\leftrightarrow)=
  \begin{cases}
    \frac{1}{3}\quad &\mbox{for}~n=3,\\
    \frac{1}{n-1}\quad &\mbox{for}~n\ge4.
  \end{cases}
  \label{eq:WSpec}
\end{equation}

Further, the LOCC verification strategy in Eq.~\eqref{eq:WStrategy} also
inspires an efficient local verification strategy without communication.
The basic idea is to replace the LOCC test $\Omega_{i,j}^{\rightarrow}$
with two local tests, performed randomly with equal probability. In the test
\begin{equation}
  \cZ^1=\sum_{i=1}^{n}\ket{0}\bra{0}^{\otimes(i-1)}\otimes
  \ket{1}\bra{1}\otimes\ket{0}\bra{0}^{\otimes(n-i)},
\end{equation}
measurement $Z^{\otimes n}$ is performed; the test is passed if the excitation is
detected exactly once.  In the test
\begin{equation}
  \Omega_{i,j}
  =\bar{\cZ}_{i,j}^1(\I\I)_{ij}
  +\bar{\cZ}_{i,j}^0(XX)_{i,j}^{+},
\end{equation}
measurement $X\otimes X$ is performed on the $i$-th and $j$-th qubit, and
measurement  $Z^{\otimes(n-2)}$ is performed on the other $n-2$ qubits; the
test is passed if one excitation is detected for measurement
$Z^{\otimes(n-2)}$, or no excitation is detected and the outcomes for the
$i$-th and $j$-th qubits coincide; see Fig.~\ref{fig:vw}. The resulting local
verification operator reads
\begin{equation}
  \Omega
  =\frac{1}{2}\cZ^1+\frac{1}{n(n-1)}\sum_{i<j}\Omega_{i,j},
  \label{eq:WLStrategy}
\end{equation}
and the spectral gap reads
\begin{equation}
  \nu(\Omega)=
  \begin{cases}
    \frac{1}{4}\quad &\mbox{for}~n=3,\\
    \frac{1}{2(n-1)}\quad &\mbox{for}~n\ge4,
  \end{cases}
  \label{eq:WLSpec}
\end{equation}
which is at most two times worse than the corresponding LOCC strategy.
Remarkably, any one-round LOCC verification strategy can be transformed to
a local strategy, in which the efficiency loss depends on the so-called branch
number \cite{Liu.etal2019}.

The verification strategy in Eq.~\eqref{eq:WStrategy} can be directly
generalized for verifying Dicke states \cite{Dicke1954}
\begin{equation}\label{eq:Dicke}
  \ket{D_n^k}=\binom{n}{k}^{-\frac{1}{2}}\sum_j
  \cP_j\Bigl\{\ket{1}^{\otimes k}\otimes\ket{0}^{\otimes (n-k)}\Bigr\},
\end{equation}
where $\sum_j\cP_j\{\cdot\}$ denotes the sum over all possible
permutations. Remarkably, the verification efficiency for Dicke states ($k\ne
0~\text{or}~n$) is equal to that for $W$ states, which is independent of the
number of excitations $k$. This makes the protocol much more efficient than the
previously known methods.

In Ref.~\cite{Li.etal2021}, Li \textit{et al.} proposed an alternative LOCC
verification strategy for $W$ states, which consists of measuring
$Z^{\otimes(n-1)}$ or $X^{\otimes(n-1)}$ on the first $n-1$ qubits and
verifying the post-measurement state on the $n$-th qubit. An advantage of this
strategy is that when $n\gg 1$
\begin{equation}
  \nu(\Omega^\rightarrow)\approx
  \begin{cases}
    \frac{0.342}{\sqrt{n}}\quad & \text{for $n$ is odd},\\
    \frac{0.287}{\sqrt{n}}\quad & \text{for $n$ is even},
  \end{cases}
\end{equation}
which is of $\cO(1/\sqrt{n})$ and asymptotically better than
Eq.~\eqref{eq:WSpec}.  The efficiency can also be further improved by about
four times when some symmetrization process is employed \cite{Li.etal2021}.

\section{Generalizations and related protocols}\label{sec:generalizations}

\subsection{The adversarial scenario}

\begin{figure}[t]
  \centering
  \includegraphics[width=.45\textwidth]{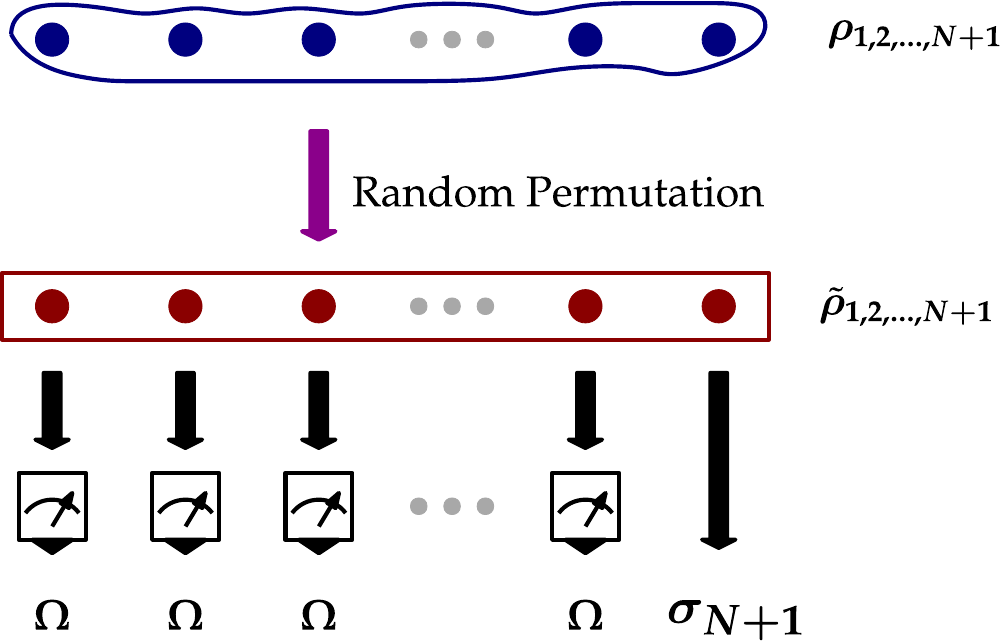}
  \caption{%
    In an adversarial scenario, the device is controlled by a potentially
    malicious adversary and can produce an arbitrarily correlated or even
    entangled state $\rho_{1,2,\dots,N+1}$ on $\cH^{\otimes(N+1)}$. The
    verifier first performs a random permutation to make the state
    permutation-invariant, i.e.,
    $\tilde{\rho}_{1,2,\dots,N+1}=\frac{1}{(N+1)!}\sum_{\pi\in
    S_{N+1}}\rho_{\pi(1,2,\dots,N+1)}$, where $S_{N+1}$ is the symmetric group.
    Then, the QSV strategy $\Omega$ is performed on each of the first $N$
    subsystems and the goal is to estimate the fidelity of the post-measurement
    state on the $(N+1)$-th subsystem $\sigma_{N+1}$, given that all the first
    $N$ subsystems pass the test.
  }
  \label{fig:adversarial}
\end{figure}

In the previous sections, the states $\sigma_1,\sigma_2,\dots,\sigma_N$
generated by the quantum devices are always assumed to be independent. In
Refs.~\cite{Zhu.Hayashi2019b,Zhu.Hayashi2019c}, Zhu and Hayashi proposed an
adversarial scenario, which makes QSV also applicable to the case of 
nonindependent sources.

In the adversarial scenario, the device is controlled by a potentially
malicious adversary and can produce an arbitrarily correlated or even entangled
state $\rho$ on $\cH^{\otimes(N+1)}$, as illustrated in
Fig.~\ref{fig:adversarial}. To verify the state produced in a certain run, the
verifier randomly chooses $N$ subsystems from $\cH^{\otimes(N+1)}$ to perform
the tests. This is also equivalent to assume that $\rho$ is
permutation-invariant on $\cH^{\otimes(N+1)}$ and the first $N$ subsystems are
chosen. For convenience, we will employ the latter description in the
following.  Now, the QSV strategy $\Omega$ is performed on each of the first
$N$ subsystems and the goal is to estimate the fidelity of the (averaged)
post-measurement state on the $(N+1)$-th subsystem, given that all the first
$N$ subsystems pass the test. That is the fidelity of the state
\begin{equation}
  \sigma_{N+1}=\frac{1}{p_\rho}\Tr_{1,2,\dots,N}
  \Bigl[\Bigl(\Omega^{\otimes N}\otimes\I\Bigr)\rho\Bigr],
  \label{eq:AdvPost}
\end{equation}
with respect to the target state $\ket{\psi}$, where
\begin{equation}
  p_\rho=\Tr\Bigl[\Bigl(\Omega^{\otimes N}\otimes\I\Bigr)\rho\Bigr]
  \label{eq:prho}
\end{equation}
is the probability that the tests are passed.
Then, the figure of merit is defined as
\begin{equation}
  F(N,\delta,\Omega):=\min_\rho\bigl\{
    \bra{\psi}\sigma_{N+1}\ket{\psi}\bmid
  p_\rho\ge\delta\bigr\},
  \label{eq:AdvF}
\end{equation}
which represents the minimum fidelity with the confidence $1-\delta$. More
precisely, the null hypothesis,
$\bra{\psi}\sigma_{N+1}\ket{\psi}<F(N,\delta,\Omega)$, is rejected with the
confidence $1-\delta$, given that all the first $N$ subsystems pass the test.
Correspondingly, to achieve a confidence $1-\delta$, the number of tests needed
is
\begin{equation}
  N(\varepsilon,\delta,\Omega)
  =\min\bigl\{N\ge 1\bmid F(N,\delta,\Omega)\ge 1-\varepsilon\bigr\}.
  \label{eq:AdvN}
\end{equation}

In the adversarial scenario, the so-called homogeneous strategy plays a crucial
role, as it is the most efficient among all verification strategies with
a given spectral gap. A verification strategy $\Omega$ is called homogeneous if
it is of the form
\begin{equation}
  \Omega=\ket{\psi}\bra{\psi}+\lambda\bigl(\I-\ket{\psi}\bra{\psi}\bigr),
  \label{eq:homogeneous}
\end{equation}
where $\ket{\psi}$ is the target state. The spectral gap is
$\nu(\Omega)=1-\lambda$ for the homogeneous strategy. The rigorous analysis of
$F(N,\delta,\Omega)$ and $N(\varepsilon,\delta,\Omega)$ is complicated even
for the homogeneous strategies. An important difference between the adversarial
and nonadversarial scenario is that the optimal strategy is not achieved when
$\lambda=0$. In particular, in the high precision limit ($\varepsilon,\delta\to
0$),
\begin{equation}
  N(\varepsilon,\delta,\Omega)\approx
  \Bigl(\lambda\ln\lambda^{-1}\Bigr)^{-1}\varepsilon^{-1}\ln\delta^{-1},
  \label{eq:AdvNLimit}
\end{equation}
which is of the same scaling with respect to $\varepsilon$ and $\delta$ as the
non-adversarial scenario in Eq.~\eqref{eq:samplingComplexity}. However, the
minimum number of tests needed is achieved when $\lambda=1/\me$.

For the general verification strategy $\Omega$, the efficiency of the
adversarial scenario depends not only on the second largest eigenvalue
$\lambda$, but also the smallest eigenvalue $\tau$ of $\Omega$. In the high
precision limit ($\varepsilon,\delta\to 0$),
\begin{equation}
  N(\varepsilon,\delta,\Omega)\approx
  h\varepsilon^{-1}\ln\delta^{-1},
  \label{eq:AdvNLimitGeneral}
\end{equation}
where the overhead $h$ reads
\begin{equation}
  h=\max\biggl\{\Bigl(\lambda\ln\lambda^{-1}\Bigr)^{-1},~
  \Bigl(\tau\ln\tau^{-1}\Bigr)^{-1}\biggr\}.
  \label{eq:AdvNLimitGeneralh}
\end{equation}
When the smallest eigenvalue $\tau$ is equal or close to zero, the adversarial
QSV is not efficient in general.  Thus, additional methods are proposed to
improve the efficiency by adding the trivial test
\cite{Zhu.Hayashi2019b,Zhu.Hayashi2019c} or other extra terms
\cite{Li.etal2019} to $\Omega$.

At last, we would like to note that for the adversarial scenario only
the unit frequency, i.e., all the previous $N$ states pass the test, has been
considered. How to generalize the problem to the non-unit frequency, like
in the case of QFE, is still an open problem. This is also of vital importance
for the practical applications of the adversarial scenario.

\subsection{Quantum process verification}

\begin{figure}[t]
  \centering
  \includegraphics[width=.45\textwidth]{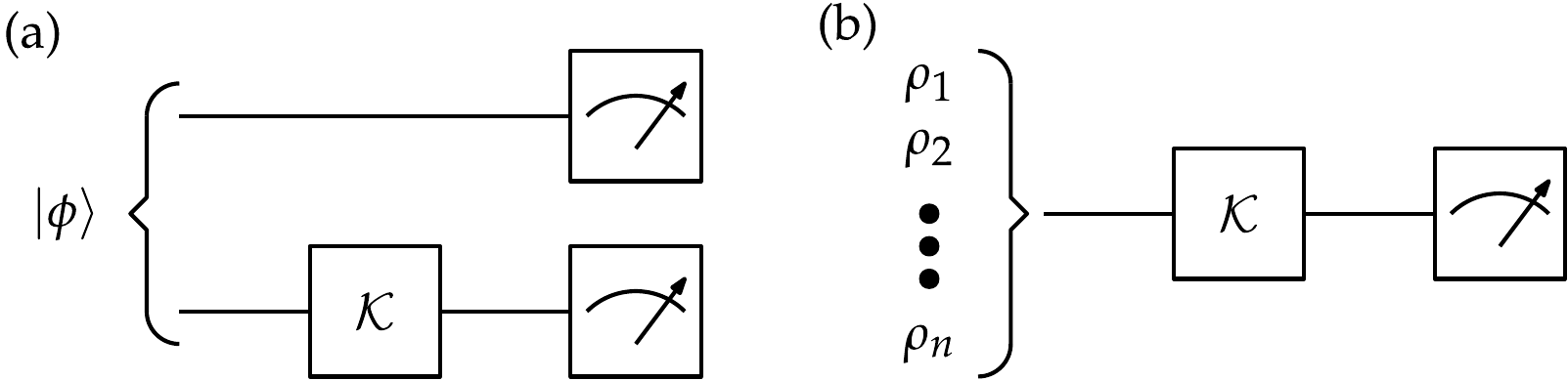}
  \caption{%
    There are two methods for verifying the quantum process $\cK$.
    (a) Ancilla-assisted QPV: by taking advantage of the Choi-Jamio{\l}kowski
    isomorphism, the verification of $K$ is transformed to the verification
    of the Choi state $\rho_{\cK}$;
    (b) Prepare-and-measure QPV: the verifier randomly chooses an input state
    $\rho_\ell$ and tests the output state with a corresponding measurement
    $\{N_\ell,\I-N_\ell\}$.
  }
  \label{fig:process}
\end{figure}

There are two main strategies for quantum process verification
(QPV); see Fig.~\ref{fig:process}. The first one is based on the
Choi-Jamio{\l}kowski isomorphism \cite{Choi1975,Jamiolkowski1972,dePillis1967}
between processes and states, which allows to relate QPV with QSV. The second
strategy is a prepare-and-measure
scheme, where certain input states are subjected to the process and
then verified.

We start our discussion with the first class of strategies \cite{Liu.etal2020,Zhu.Zhang2020,Zeng.etal2020}. Consider
the (unnormalized) maximally entangled bipartite state
$\ket{\phi}=\sum_{\alpha=1}^{d}\ket{\alpha}_A\ket{\alpha}_S$
between a quantum system $\cH_S$ and an ancilla system $\cH_A$.
The Choi-Jamio{\l}kowski isomorphism $J$ is defined as
\begin{equation}
  J(\cE):=\id\otimes\cE\bigl(\ket{\phi}\bra{\phi}\bigr)=
  \sum_{\alpha,\beta=1}^{d}\ket{\alpha}\bra{\beta}
  \otimes\cE\bigl(\ket{\alpha}\bra{\beta}\bigr),
  \label{eq:ChoiMat}
\end{equation}
where $\cE:\cB(\dC^d)\to\cB(\dC^d)$ is a map (quantum process) on the system
$\cH_S$ only, and $J(\cE)\in\cB(\dC^d\otimes\dC^d)$ is usually called the Choi
matrix of the process $\cE$.  Conversely, $\cE$ can also be obtained from
$J(\cE)$ as
\begin{equation}
  \cE(\rho)=\Tr_{A}\Bigl[\Bigl(\rho^T\otimes\I\Bigr)J(\cE)\Bigr].
  \label{eq:ChoiRep}
\end{equation}

Tomographically, Eq.~\eqref{eq:ChoiRep} implies that once the Choi
matrix $J(\cE)$ is determined, all the information of the process $\cE$ is
known. Thus, one can verify a quantum process $\cK$ indirectly by
verifying the corresponding Choi state
\begin{equation}
 \rho_{\cK}:=\frac{J(\cK)}{\Tr[J(\cK)]}
  \label{eq:ChoiState}
\end{equation}
instead. Especially, when $\rho_{\cK}$ is pure (e.g., if $\cE$ is unitary),
one can apply the QSV protocols for verifying $\cK$.

For simplicity, we only consider the case when $\cK$ is a unitary gate $U$,
i.e., the verification of quantum gates or quantum circuits. In general, $\cE$
is not necessary to be trace-preserving, which makes it possible to deal with
post-selection or particle losses \cite{Liu.etal2020}. Suppose that a quantum
device is promised to perform a unitary gate $U$, but a quantum process $\cE$
is performed in practice. We want to use the hypothesis testing method to
verify this claim with high confidence. The entanglement gate fidelity
\begin{equation}
  F_e(\cE, U):=F(\rho_{\cE},\rho_{U})=\Tr(\rho_{\cE}\rho_{U})
  \label{eq:fidelity}
\end{equation}
is employed as a benchmark, which is only different from the average gate
fidelity by an affine function \cite{Horodecki.etal1999}.

Due to the Choi-Jamio{\l}kowski isomorphism, one can transform the verification
of $U$ to the verification of the pure Choi state
$\rho_U=\ket{\psi_U}\bra{\psi_U}$, where
\begin{equation}
  \ket{\psi_U}=\frac{1}{\sqrt{d}}\sum_{\alpha=1}^d
  \ket{\alpha}\otimes U\ket{\alpha}.
\end{equation}
This method is called the ancilla-assisted QPV. As $\rho_U$ is maximally
entangled, one can verify $\rho_U$ with the spectral gap $\nu(\Omega)=d/(d+1)$
as shown in Eq.~\eqref{eq:MESSpec}.  In actual experiments, one usually has
more restrictions to the allowed measurements, e.g., each party should be
measured locally for verifying the multi-qubit gates. In general, the
worst-case failure probability in each run is always bounded by
\begin{equation}
  \max_{F(\rho_\cE,\rho_U)\le 1-\varepsilon}\Tr(\Omega\rho_\cE)
  \le 1-\varepsilon\nu(\Omega).
\end{equation}
Correspondingly, the confidence $1-\delta$ is still bounded by
Eq.~\eqref{eq:failureProbability} or Eq.~\eqref{eq:CHbound}.

The second class of strategies is the prepare-and-measure QPV, which does not
require an additional ancilla system or maximally entangled input states.  In
the prepare-and-measure QPV, the verifier randomly chooses an input state
$\rho_\ell$ with probability $p_\ell$ and test the output state with the
measurement $\{N_\ell, \I-N_\ell\}$.  If the measurement outcome is $N_\ell$,
then we say the channel $\cE$ passes the test; otherwise we say $\cE$ fails the
test.  Similar to QSV, we require that the target gate $U$ always passes the
test, i.e.,
\begin{equation}
  \Tr(U\rho_\ell U^\dagger N_\ell)=1.
  \label{eq:UPass}
\end{equation}
For convenience, we denote the prepare-and-measure strategy as
\begin{equation}
  \Xi=\sum_\ell p_\ell\rho_\ell^T\otimes N_\ell.
  \label{eq:PMPV}
\end{equation}
Then in each run the worst-case failure probability is given by
\begin{equation}
  \max_{F_e(\cE,U)\le 1-\varepsilon}\sum_\ell
  p_\ell\Tr[\cE(\rho_\ell)N_\ell]
  =\max_{F_e(\cE,U)\le 1-\varepsilon}\Tr[\Xi J(\cE)].
  \label{eq:efficiencyQPV}
\end{equation}

A remarkable result on QPV is that every one-way LOCC QSV strategy for the Choi
state $\rho_U$ can be transformed to a prepare-and-measure QPV strategy
\cite{Liu.etal2020}.  According to Eq.~\eqref{eq:oneWayStrategy}, the one-way
adaptive QSV strategy takes on the general form
\begin{eqnarray}
  \Omega^\rightarrow=\sum_\ell M_\ell \otimes N_\ell,
  \label{eq:oneWayQSV}
\end{eqnarray}
such that $\{M_\ell\}_\ell$ is a POVM on the ancilla system $\cH_A$ and
$\{N_\ell, \I-N_\ell\}$ is a pass or fail test on the system $\cH_S$ which
depends on the measurement outcome of $\{M_\ell\}_\ell$. Now,
$\Omega^\rightarrow$ can be converted to a prepare-and-measure QPV strategy of
the form in Eq.~\eqref{eq:PMPV} by letting
\begin{equation}
  p_\ell=\frac{\Tr(M_\ell)}{d},~\rho_\ell
  =\frac{M_\ell^T}{\Tr(M_\ell)},~N_\ell=N_\ell,
  \label{eq:oneWayInput}
\end{equation}
and the failure probability defined in Eq.~\eqref{eq:efficiencyQPV} is bounded
by
\begin{equation}
  \max_{F_e(\cE,U)\le 1-\varepsilon}\Tr[\Xi J(\cE)]
  \le 1-\varepsilon\nu(\Omega^\rightarrow).
  \label{eq:errorQPV}
\end{equation}
Again, the confidence $1-\delta$ is bounded by
Eq.~\eqref{eq:failureProbability} or Eq.~\eqref{eq:CHbound}. By taking
advantage of the corresponding QSV strategies in Sec.~\ref{sec:qsv}, many
widely-used quantum gates can be efficiently verified, including general single
qubit and qudit gates, Clifford gates, $C^{(n-1)}Z$ and $C^{(n-1)}X$ gates, as
well as CSWAP gates \cite{Liu.etal2020,Zhu.Zhang2020,Zeng.etal2020}. The
adversarial scenario for quantum gate verification was also considered in
Ref.~\cite{Zeng.etal2020}.

Using photonic platforms, the first experimental verification of quantum gates 
including a two-qubit controlled-not gate and a three-qubit Toffoli gate using 
only local state preparations and measurements was reported in 
Ref.~\cite{Zhang.etal2022}.  The experimental results show that, by using only 
1600 and 2600 measurements on average, a $95\%$ confidence can be drawn that 
the implemented controlled-not gate and Toffoli gate have fidelities at least 
$99\%$ and $97\%$, respectively.  This is substantially more efficient than 
quantum process tomography, thus successfully demonstrated the superior low 
sample complexity and experimental feasibility of quantum process verification.  
See also Ref.~\cite{Luo.etal2022} for a proof-of-principle optical 
demonstration of quantum gate verification of two general single-qubit gates.

\subsection{Quantum entanglement verification}
The QSV protocol, more precisely, the QFE protocol is also closely related to
the statistical entanglement verification method proposed by Dimi\'c and
Daki\'c and by Saggio \textit{et al.} in
Refs.~\cite{Dimic.Dakic2018,Saggio.etal2019}; see also the very recent review 
paper \cite{Morris.etal2021}. In their method, the verifier
also perform a set of random pass or fail tests
$\{\Omega_\ell,\I-\Omega_\ell\}$, such that no separable state can pass the
test with probability higher than $q_s$. Similar to Eq.~\eqref{eq:CHbound},
if in actual experiments the frequency of pass instances $f$ is larger than
$q_s$, the verifier can conclude the state is entangled with the confidence
$1-\delta$, where
\begin{equation}
  \delta\le\me^{-D[f\|q_s]N},
  \label{eq:CHboundSep}
\end{equation}
and $D[\cdot\|\cdot]$ is the Kullback–Leibler divergence.

To see the connection to QFE, let us consider the fidelity-based entanglement
witness \cite{Guehne.Toth2009} for bipartite states. This is a limited class
of witness operators \cite{Weilenmann.etal2020,Guehne.etal2021} but widely used
in real experiments. According to Ref.~\cite{Guehne.etal2021}, we can just
consider the witness operator
\begin{equation}
  W=\frac{1}{d}\I\otimes\I-\ket{\psi}\bra{\psi},
  \label{eq:maxWitness}
\end{equation}
where $\ket{\psi}$ is some maximally entangled state. This implies that for
a quantum state $\sigma$
\begin{equation}
  \sigma\in\Sep\quad\Rightarrow\quad
  \bra{\psi}\sigma\ket{\psi}\le\frac{1}{d},
  \label{eq:FEW}
\end{equation}
which essentially transforms an entanglement witness problem to a fidelity
estimation problem. Now, suppose that we use the optimal QSV strategy $\Omega$ in
Eq.~\eqref{eq:MESStrategyCont} or \eqref{eq:MESStrategyDisc} for the maximally
entangled state, where the spectral gap is $\nu(\Omega)=d/(d+1)$.  Then,
Eqs.~(\ref{eq:worstcase},\,\ref{eq:FEW}) imply that
\begin{equation}
  \max_{\sigma\in\Sep}\Tr(\Omega\sigma)
  \le 1-\Bigl(1-\frac{1}{d}\Bigr)\nu(\Omega)
  =\frac{2}{d+1}.
\end{equation}
Similar to Eq.~\eqref{eq:CHbound}, the null hypothesis that all the produced
states $\sigma_1,\sigma_2,\dots,\sigma_N$ are separable can be rejected with
the confidence $1-\delta$ \cite{Dimic.Dakic2018,Saggio.etal2019} with
\begin{equation}
  \delta\le\me^{-D\bigl[f\big\|\frac{2}{d+1}\bigr]N},
  \label{eq:CHboundEW}
\end{equation}
where $f$ is the frequency of the pass instances. Actually, the value
$1-\delta$ can also be interpreted as the confidence of the presence of
entanglement in the averaged state
\begin{equation}
  \bar{\sigma}=\frac{1}{N}\sum_{k=1}^N\sigma_k.
\end{equation}

The method can be directly generalized to detect (genuine) multi-partite
entanglement~\cite{Dimic.Dakic2018,Saggio.etal2019}, as the fidelity-based
entanglement witness of the form in Eq.~\eqref{eq:FEW} also exists for
multi-partite states \cite{Guehne.Toth2009}.

More generally, in Ref.~\cite{Saggio.etal2019}, Saggio \textit{et al.} also
showed that this method can in principle go beyond the fidelity-based
entanglement witness. The basic idea is that any entanglement witness can be
decomposed into local operators, which can then be measured probabilistically.
This leads to a general method for transforming any entanglement witness
operator to a statistical entanglement verification strategy.

Then, the experimental verification of entanglement in a photonic 
six-qubit cluster state was presented in Ref.~\cite{Saggio.etal2019}.
It showed that the presence of entanglement can be certified with at 
least $99.74\%$ confidence by using 20 copies of the quantum state.
Additionally, genuine six-qubit entanglement can be verified with at 
least $99\%$ confidence by using 112 copies of the state. These 
results make it possible to apply the method to verify large-scale quantum 
devices.

\subsection{Emerging research directions}
Apart from the aforementioned generalizations there are also other
emerging research directions that further extend the applicability
and efficiency of QSV protocols. Here, we mention only a few of them.

As proved in Ref.~\cite{Pallister.etal2018}, the QSV strategy satisfying
Eq.~\eqref{eq:OmegaEigen}, i.e., the zero type II error,  is optimal in the 
asymptotic limit. Roughly speaking, this is because the Kullback–Leibler 
divergence satisfies that
\begin{equation}
  D(f\|f-\varepsilon)=
  \begin{cases}
    \cO(\varepsilon) \quad& f=1,\\
    \cO(\varepsilon^2) \quad& 0<f<1,
  \end{cases}
\end{equation}
when $\varepsilon\to0$, and Eq.~\eqref{eq:OmegaEigen} is necessary to ensure
that the frequency of pass instances is $f=1$. This 
advantage, however,  is not robust as it exists only 
in the idealized situation when the fidelity of the 
produced states is exactly $1$. Thus, in practice, it is
interesting to study whether the efficiency of the QFE protocol can be improved
by relaxing the constraint in Eq.~\eqref{eq:OmegaEigen}.
In Ref.~\cite{Thinh.etal2020}, the authors investigate a related problem under
the QSV framework, i.e., with the hypotheses in
Eqs.~(\ref{eq:QSValt},~\ref{eq:QSVnull})
\footnote{Note that the choices of null and alternative hypotheses are the
opposite in Ref.~\cite{Thinh.etal2020}, with Eq.~\eqref{eq:QSValt} being the null
hypothesis and  Eq.~\eqref{eq:QSVnull} being the alternative hypothesis.}.
Instead of choosing zero Type II error $\bra{\psi}\Omega\ket{\psi}=1$, the
authors consider the bounded Type II error,
\begin{equation}
    \bra{\psi}\Omega\ket{\psi}=1-\beta_0,
\end{equation}
under which the minimization of type I error is studied.

In Ref.~\cite{Liu.etal2021}, an alternative implementation of the QSV protocol 
is proposed by using so-called quantum nondemolition (QND) measurements, which 
are the type of measurements that leave the post-measurement quantum states 
undestroyed, thus allowing repeated or sequential measurements.  The protocol 
fully explores the use of sequentially constructed QND measurements for state 
verification instead of the probabilistic construction in standard QSV 
strategies.  Under such a design, not only the target states can be preserved, 
but also the protocol turns to be equivalent to the optimal global strategy in 
terms of the verification efficiency.  Moreover, the protocol is robust in the 
sense that the order of the sequential measurements can be arbitrarily 
constructed which is rather friendly to experimental implementations.  Very 
recently, Miguel-Ramiro \textit{et al.} showed with collective local 
measurements, one can also significantly reduce the destruction of the verified 
states \cite{MiguelRamiro.etal2022}.

In Ref.~\cite{Gocanin.etal2022}, the authors generalize QSV to the
device-independent scenario. Their method is based on the self-testing
properties of entangled quantum states \cite{Mayers.Yao2004,Supic.Bowles2020}.
It is shown that a quantum state can be device-independently verified if there
exists a robust self-testing protocol for it. Moreover, a systematic method is
proposed to construct device-independent QSV protocols and the confidence
is derived from the robustness of the corresponding self-testing protocol.
In the same paper, the authors also consider the device-independent
adversarial scenario. Compared with the device-dependent adversarial scenario
in Sec.~\ref{sec:generalizations}, their method has the advantage that
more than one remaining state can be certified, while the disadvantage is
that the method is only applicable to independent copies.

\section{Conclusions and Outlooks}
\label{sec:conclusions}
The framework of hypothesis testing is powerful and allows to analyze 
experimental data in an efficient manner. The methods for quantum state 
verification described in this review article are therefore important 
to analyze current and future experiments, where only a limited amount 
of data is available. Even more, although the considered situation may 
seem artificial at first sight, the insights from optimal quantum state 
verification protocols also turn out to be useful for other tasks, 
such as fidelity estimation or entanglement verification in a realistic 
scenario. Consequently, there are several open problems where the 
presented results may be useful or may be extended: 
\begin{itemize}

\item So far, we considered only discrete systems, mainly multi-qubit 
systems. It is highly desirable to develop a similar theory also for 
continuous-variable systems or hybrid systems, where some parties are 
finite-dimensional, while others are not. While finishing this review, 
first results in this direction have been published \cite{Liu.etal2021b, 
Wu.etal2021}.

\item The current approach to QSV is designed for the verification of 
pure states. For more realistic scenarios an extension to the case of
mixed states is needed. For instance, how can one estimate mixed state
fidelities in an optimized manner? The known results show that this 
problem is complicated even if entangled measurements are 
allowed~\cite{Badescu.etal2019,Chen.etal2021}.

\item The scheme of QSV can be seen as a characterization of a quantum 
source, where one asks whether the source produces always the same 
quantum state. For characterizing sources, however, many other questions
may be asked, e.g., concerning the stability of the source or potential 
drifts \cite{Proctor.etal2020}. Developing a full theory for these effects 
from a rigorous statistical viewpoint is highly desirable.

\item In the current literature on QSV one typically considers the 
number of copies $N$ to be fixed. In a realistic experiment one 
may, however, keep it flexible, and abort an experiment if a desired
confidence has been reached. This leads to the notion of sequential
tests, which have recently been attracted some interest in the
quantum information community \cite{Milazzo.etal2019, Martinez-Vargas.etal2021}.

\item The presented protocols of QSV bear some similarity with other 
protocols for quantum information processing. For instance, the fact 
that the desired objects pass with certainty some test also arises in
entanglement distillation protocols \cite{Duer.Briegel2007} and Hardy-type tests
of non-locality \cite{Hardy1992}. It may be fruitful to formalize this similarity.

\item Finally, recently the notion of quantum algorithmic measurements 
(QUALMs) has been introduced \cite{Aharonov.etal2021}, where the measuring 
party has coherent access to a quantum computer. This extended notion of 
measurements has the potential to lead to radically novel and enhanced 
strategies for QSV.
\end{itemize}

In summary, we believe that the topic of quantum state verification is just
emerging and is likely to have a fruitful impact on other topics in quantum
information processing in the future.

\begin{acknowledgments}
We thank Geng Chen, Rui Han, Yun-Guang Han,  Masahito Hayashi, Qiongyi He, Zhibo 
Hou, Martin Kliesch, Yinfei Li, Zihao Li, Ye-Chao Liu, Ramon Mu\~noz Tapia, 
Gael Sent\'{\i}s, G\'eza T\'oth, Kun Wang, Guo-Yong Xiang, Rui-Qi Zhang, 
Xiangdong Zhang, Huangjun Zhu for discussions on quantum state verification.

This work was supported by the Deutsche Forschungsgemeinschaft (DFG, German 
Research Foundation, project numbers 447948357 and 440958198), the Sino-German 
Center for Research Promotion (Project M-0294), and the ERC (Consolidator Grant 
683107/TempoQ). J.S. acknowledges support by the National Natural Science 
Foundation of China (Grants No. 12175014 and No. 11805010).
\end{acknowledgments}

\bibliography{QuantumInf}

\end{document}